\documentclass[preprint2]{aastex}
\usepackage{amssymb,amsmath}
\usepackage{graphicx, natbib}
\usepackage{longtable, float, subfigure}
\usepackage[T1]{fontenc}
\usepackage[latin1]{inputenc}
\def\kms{km s$^{-1}$}
\def\et{{et~al.}}

\def\arcmin{\ifmmode^\prime\;\else$^\prime$\fi}
\def\arcsec{\ifmmode^{\prime\prime}\;\else$^{\prime\prime}$\fi}
\def\deg{\ifmmode^\circ\;\else$^\circ$\fi}

\def\jybm{Jy~bm$^{-1}$}
\def\hi{\ion{H}{1}}
\def\hdeg{{0$\fdg5$}~\ion{H}{1}~cloud}

\shorttitle{Extended \hi\ Structure Around NGC 1569}
\shortauthors{Johnson}

\begin{document}
\title{Determining the Nature of the Extended \hi\ Structure Around LITTLE THINGS Dwarf Galaxy NGC 1569}
\author {Megan Johnson\altaffilmark{1,2}
}

\altaffiltext{1}{Lowell Observatory, 1400 West Mars Hill Road, Flagstaff, AZ 86001; dah@lowell.edu}
\altaffiltext{2}{National Radio Astronomy Observatory, PO Box 2, Green Bank, WV 24915; mjohnson@nrao.edu}

\begin{abstract}
This work presents an extended, neutral Hydrogen emission map around Magellanic-type dwarf irregular galaxy (dIm) NGC 1569.   
In the Spring of 2010, the Robert C. Byrd Green Bank Telescope (GBT) was used to map a 9$\arcdeg$ x 2$\arcdeg$ region in \hi\ line emission that includes NGC 1569 and IC 342 as well as two other dwarf galaxies. 
The primary objective for these observations was to search for structures potentially connecting NGC 1569 with IC 342 group members in order to trace previous interactions and thus, provide an explanation for the starburst and peculiar kinematics prevalent in NGC 1569. 
A large, half-degree diameter \ion{H}{1} cloud was detected that shares the same position and velocity as NGC 1569.
Also, two long structures were discovered that are reminiscent of intergalactic filaments extending out in a v-shaped manner from NGC 1569 toward UGCA 92, a nearby dwarf galaxy.  These filamentary structures extend for about 1$\fdg$5, which is 77 kpc at NGC 1569.  There is a continuous velocity succession with the \hdeg, filaments, and main body of the galaxy.  The \hdeg\ and filamentary structures may be foreground Milky Way, but are suggestive as possible remnants of an interaction between NGC 1569 and UGCA 92.  The data also show two tidal tails extending from UGCA 86 and IC 342, respectively.  These structures may be part of a continuous \hi\ bridge but more data are needed to determine if this is the case.
\end{abstract}
\keywords{galaxies: individual (NGC 1569, UGCA 92, UGCA 86, IC 342) --- galaxies: dwarf galaxies --- galaxies: starburst}

\section{Introduction}\label{intro}
In the spring of 2010, the National Radio Astronomy Observatory's (NRAO\footnote{The National Radio Astronomy Observatory is a facility of the National Science Foundation operated under cooperative agreement by Associated Universities, Inc.}) Robert C. Byrd Green Bank Telescope (GBT) was used to map a large region around the LITTLE (Local Irregulars That Trace Luminosity Extremes) THINGS (The HI Nearby Galaxy Survey) Magellanic-type dwarf irregular (dIm) galaxy NGC 1569 in \ion{H}{1} line emission.  LITTLE THINGS is an international collaboration formed to study star formation in gas rich dwarf galaxies.  There are 41 dwarfs in our sample and we have a rich multi-wavelength data set including high spatial ($\sim$6$\arcsec$) and spectral (2.6 or 5.2 \kms) resolution \hi\ data acquired with the NRAO Very Large Array (VLA).  We are adding to this multi-wavelength dataset wide-field GBT \hi\ emission maps around each object.  The purpose of these observations is to study the extended \hi\ environment around these mostly isolated, field dwarf galaxies.  NGC 1569 is one of the galaxies in the LITTLE THINGS survey and a detailed study of the high resolution VLA data \citep[][hereafter Paper I]{joh12} prompted a follow-up investigation of high sensitivity, deep \hi\ GBT observations.  Please see \citet{hun12} for more information on the LITTLE THINGS survey. 

The GBT was used to search for extended neutral hydrogen structures that would give evidence of tidal interactions between NGC 1569 and other IC 342 group members.  Dwarf galaxies are believed to be the building blocks of larger galaxies \citep{bel03, col00}, yet the role of dwarf-dwarf galaxy interactions and mergers in the local universe is not well known.  Recently, however, there have been attempts from theory to explain some of the local dwarf systems as dwarf-dwarf induced mergers or interactions \citep{bes12, yoz12, bek08, don08}.  Even still, observations are lacking for a complete statistical view of the role of dwarf-dwarf interactions and mergers, especially for relatively isolated dwarf systems.  Searching for extended \hi\ filaments, bridges, tails, clouds, rings, and other structures around likely interacting dwarf galaxy candidates offers the most probable method for identifying merging and interacting systems as \hi\ is usually the most massive tidal component in a gas-rich interaction or merger \citep{bes12, sco12, sen09, man08, loc03}. If dwarf galaxies mimic their more massive spiral galaxy counterparts, then starburst systems are the most likely objects to have undergone recent interaction or merger activity \citep[e.g.,][]{ivi12, tad11}.

NGC 1569 is one of the nearest examples of a starburst dIm galaxy, and it contains three supermassive star clusters (SSCs).  Paper I presents a detailed study of the gas and stellar kinematics of NGC 1569 and compares the kinematics with the morphology of H$\alpha$ and optical images.  We determined a ratio of maximum rotation speed to stellar velocity dispersion, $V_{\rm max}$/$\sigma_{\rm z}$ = 2.4 $\pm$ 0.7, that is consistent with a thick stellar disk.  A region near SSC A where the stellar velocity dispersion {\it increases} by $\sim$13 \kms\ was detected, potentially indicating an area where a second generation of stars formed out of shocked, accelerated gas expanding around SSC A.  Paper I modeled the contribution of stars and gas to the gravitational potential of the galaxy and found that the stellar mass dominates the gravitational potential in the inner 1 kpc (nearly half) of the disk, which is unusual in disk galaxies and dwarfs in particular.

In Paper I, we observed that the \ion{H}{1} velocity field of NGC 1569 has strong non-circular motions over the entire disk.  These non-circular motions were deconvolved from ordered motion and a non-circular motion (NCM) \ion{H}{1} cloud was detected.  We concluded that this NCM cloud is impacting the disk southwest of the SSCs.  This cloud was also seen in the work by \citet{sti02}. The NCM cloud is likely responsible for the recent starburst.  In this picture, this cloud, with a velocity difference $\sim$50 \kms\ from the systemic velocity of NGC 1569, shocked a dense \ion{H}{1} ridge that stretched across what is now the stellar disk.  As the NCM cloud fell into the center of the galaxy, it caused the middle of the dense ridge to collapse, thus creating the most recent starburst episode during which the SSCs formed.  Paper I observed an ultra-dense \ion{H}{1} cloud that lies to the west of the SSCs that is believed to be a remnant of the dense \ion{H}{1} ridge. The NCM cloud is hitting the galaxy where the ultra-dense cloud and the stellar disk meet (see Figures 1, 18 and 19 in Paper I).

From the high resolution VLA data, Paper I found tenuous \ion{H}{1} emission extending south and north of the disk of NGC 1569.  Because of the recent starburst, the presence of the NCM and ultra-dense \ion{H}{1} clouds, peculiar stellar and gas kinematics, and tenuous \ion{H}{1} emission around the disk of the galaxy, 
Paper I determined that NGC 1569 is likely undergoing an interaction or merger.

The triggering mechanism for the starburst, however, has remained controversial.  Until recently, NGC 1569 was thought to be an isolated system in the projected direction of the IC 342 group of galaxies, thus making the interaction scenario unlikely.  It lies close to the Galactic plane and is highly obscured, which made accurate distance determination difficult.  \citet{hun82} found a distance of 4.7 Mpc from the center of the \hi-line velocity profile where they assumed a Hubble constant of 50 \kms.  \citet{kar94} produced a distance of 1.8 Mpc from fitting a Johnson's $B-V$ vs $V$ color magnitude diagram.  The most commonly quoted distance of late was 2.2 $\pm$ 0.6 Mpc, which came from \citet{isr88} who constrained the foreground extinction using UV photometry, thus modifying the previously adopted distance moduli from \citet{abl71} and \citet{arp85}.
\citet{gro08} detected the tip of the red giant branch using the {\it Hubble Space Telescope} ({\it HST})
and published a new distance of 3.36 $\pm$ 0.20 Mpc placing NGC 1569 well within the IC 342 galaxy group.  This distance has been refined to the most up-to-date published distance of 2.96 $\pm$ 0.22 Mpc \citep{gro12}, which puts NGC 1569 at the near edge of the IC 342 galaxy group.  This is the distance assumed for NGC 1569 in this work.

Having a galaxy group association opens up the possibility that NGC 1569 may have undergone an interaction or merger, which is a natural way to produce a starburst and explain the observed chaotic kinematics of the stars and gas.  However, \citet{bro04} find that not all interactions form starbursts in dwarf galaxies and they suggest that gas mass oscillations within a dark matter halo can be a dominant triggering mechanism for starburst dwarfs.  Detecting \hi\ tidal structures in the outskirts of NGC 1569 can be valuable for identifying the starburst triggering mechanism.

This paper is structured as follows: Section \ref{sec:1} outlines the observations and data reductions; Section \ref{sec:5} gives the results; Section \ref{sec:2} examines the analysis and presents a discussion; Section \ref{sec:3} explores the implications of the extended \ion{H}{1} emission; and Section \ref{sec:4} summarizes the conclusions.

\section{Observations and Data Reduction}\label{sec:1}

\begin{figure}[htbp]
\centering
\includegraphics[width=4in]{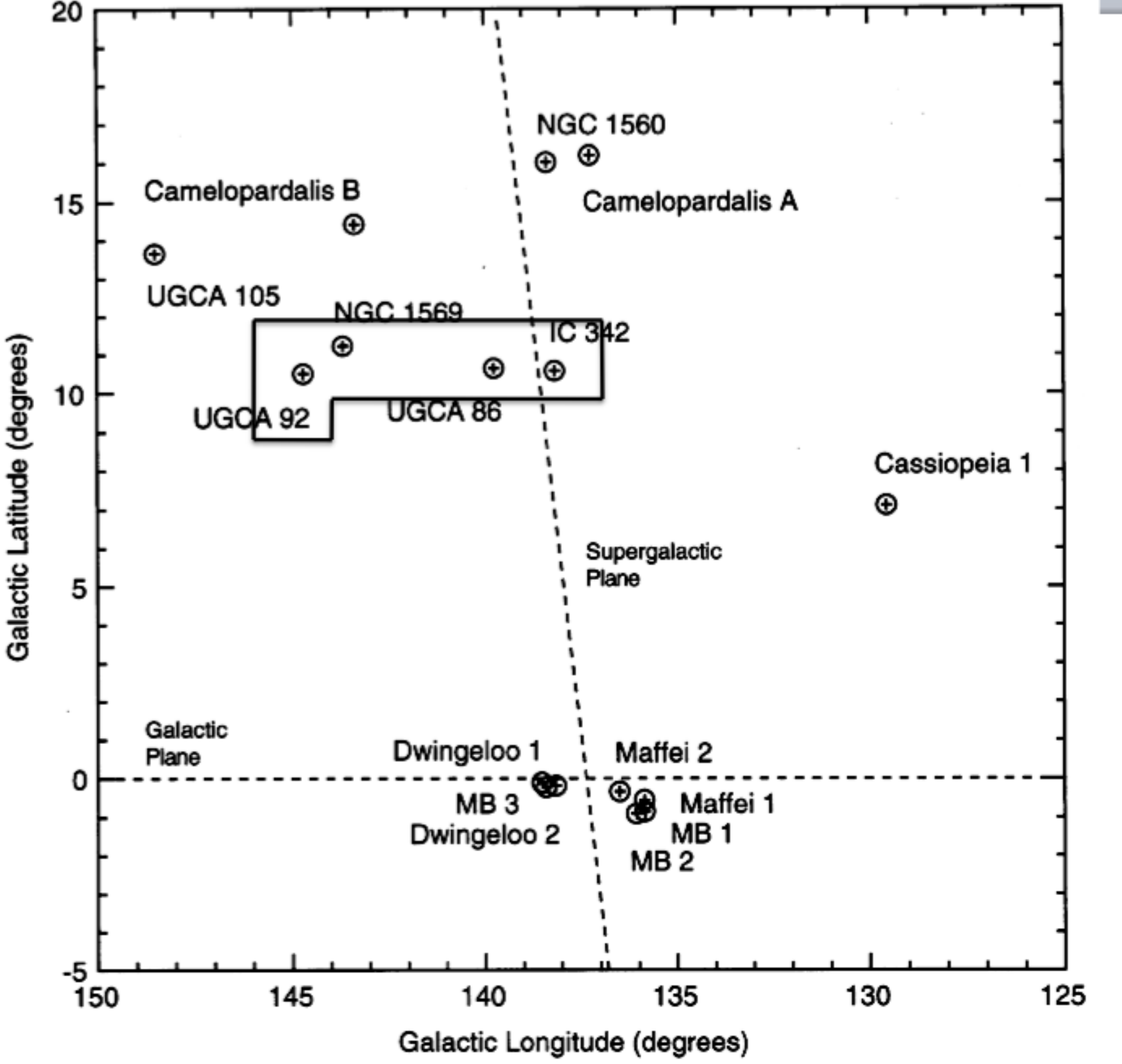}
\hfil
\includegraphics[width=4in]{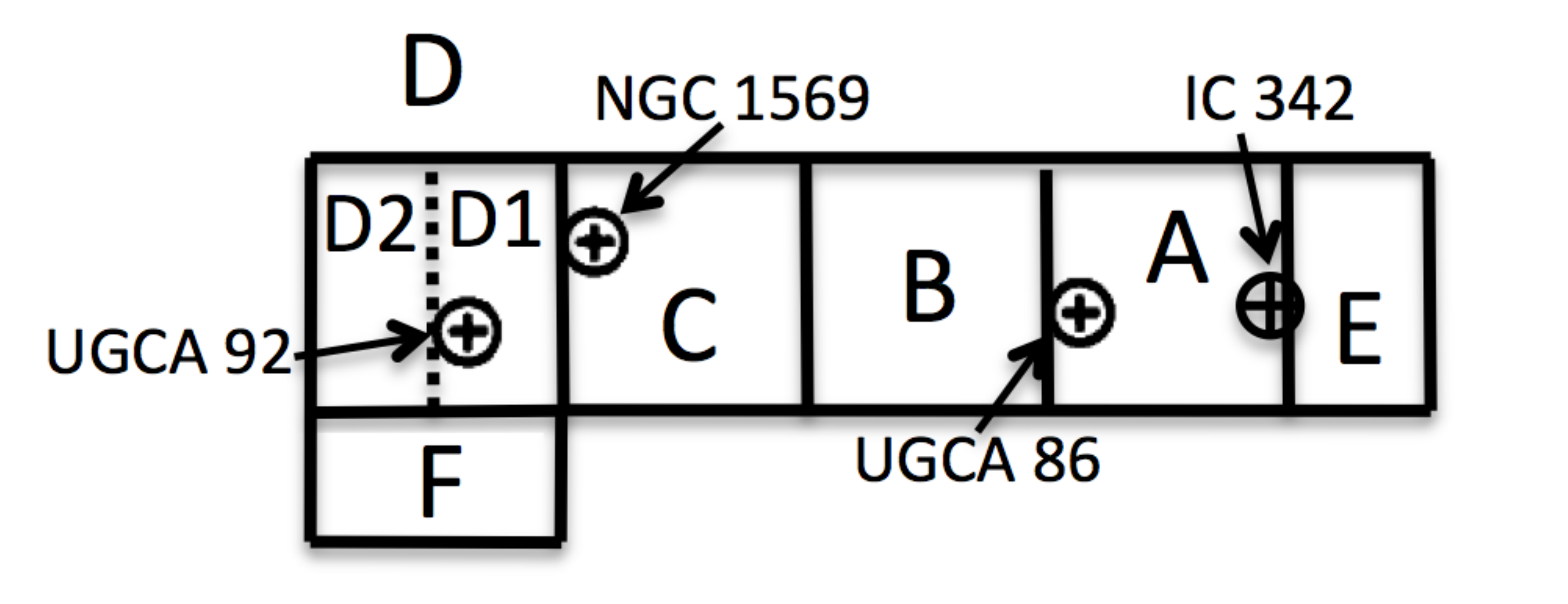}
\caption{Top: Map of IC 342/Maffei galaxy group from \citet{but99} showing the region mapped with the GBT outlined by the black rectangular region. Bottom: Total mapped region divided into the six observing areas. Region D was originally two pieces, D1 and D2, but became a single region in March.}
\label{fig:11}
\end{figure}

The GBT data were collected over the course of four months, 2010 February -- 2010 May.  Table \ref{tab:obs} lists the details of the telescope and observations.  The observations were taken using the L-band receiver centered at 1.420 GHz using a bandwidth of 12.5 MHz and 16,384 channels, producing an initial channel separation of 0.158 km s$^{-1}$.   The GBT has a beam size, half-power beam width (HPBW), of 9$\farcm$1.  On-the-fly mapping was used with Nyquist sampling, which produced a pixel size of $\sim$3$\farcm$5.  
A 5$\sigma$ (RMS noise) \ion{H}{1} sensitivity of $N_{\rm HI}$ = 1.4 x $10^{18}$ cm$^{-2}$ was achieved for a velocity line width, full-width at half-maximum (FWHM), of 20 \kms.
In order to observe the full 9$\arcdeg$ $\times$ 2$\arcdeg$ ($l$ $\times$ $b$) region shown in Figure \ref{fig:11}, the total area was divided up into four 2$\arcdeg$ $\times$ 2$\arcdeg$ squares (regions A, B, C, and D in bottom panel of Figure \ref{fig:11}) and one 1$\arcdeg$ $\times$ 2$\arcdeg$ rectangle (region E in bottom panel of Figure \ref{fig:11}).  Additionally, one 2$\arcdeg$ $\times$ 1$\arcdeg$ rectangle (region F in Figure \ref{fig:11}) was added to study an area to the southeast of NGC 1569.  Each region was mapped six times in rows of constant Galactic longitude where each row was separated by a pixel, 3$\farcm5$.  These maps were co-added together for a total integration time of 36 seconds per pixel.  Table \ref{tab:obs} lists the dates that each region was mapped.

To calibrate the spectra, in-band  
frequency switching was used and the frequency was switched between the center of the line of the rest frequency and $-3.5$ MHz from the rest frequency for a full switching cycle of 1 second.  In-band frequency switching was preferred to position switching because the Milky Way emission is very close to NGC 1569 in position and systemic velocity.  To calibrate the flux, observations of 3C147 were done at the start of most observing runs.

The data reductions were done using GBTIDL\footnote{Developed by NRAO; documentation at http://gbtidl.sourceforge.net} and AIPS\footnote{The Astronomical Image Processing System (AIPS) has been developed by the NRAO; documentation at http://www.aoc.nrao.edu/aips}.  Standard calibration procedures in GBTIDL were used to 
calibrate the data and trim the edges of the spectra.  There were a few RFI signals present in the spectra that were generated from within the GBT receiver room, which were less than one channel wide, and were removed by interpolating the surrounding channels.
Also, GBTIDL was used to remove residual instrumental baseline structure by applying a low-order polynomial.  The bandwidth was wide enough that there were more than 350 channels (after smoothing) that were free of emission to either side of the channels with emission that were used to model the baseline effectively.  The baseline stability of the GBT is unparalleled because of its unblocked aperture and its location within the National Radio Quiet Zone.  Therefore, the uncertainties from the baseline fitting are minimal.

The final data were Hanning and boxcar smoothed to a velocity resolution and channel separation of 0.81 km s$^{-1}$, leaving 1201 channels spanning a velocity range of -461.1 to 505.2 km s$^{-1}$ (all velocities in this paper are heliocentric).   See Table \ref{tab:obs} for a list of the telescope parameters.  All of the spectra were combined ({\sc dbcon}) into a single cube and gridded ({\sc sdgrd}) in AIPS using a spherical Bessel function \citep{man07}.  

The map includes four galaxies, NGC 1569, UGCA 92 (the nearest object to NGC 1569), IC 342 (the nearest large spiral galaxy to NGC 1569), and UGCA 86 (a dwarf galaxy close to IC 342 in the direction of NGC 1569). Table \ref{tab:1} gives some of the global parameters of each galaxy in the map.

\section{Results}\label{sec:5}

Figure \ref{fig:1}
shows individual channel maps highlighting the region around NGC 1569 and UGCA 92.  
The velocities are given in the lower right corner of each panel.  The intensity scale is the flux density in \jybm.
Each frame is one channel, 0.81 \kms, apart.  
NGC 1569 and UGCA 92 are marked in the first velocity panel, -116.4 \kms.  In velocity panel -114.0 \kms, a large, \hdeg\ is identified and it is visible from -115.6 to -111.6 \kms, which are
velocities coincident with those of NGC 1569.  Velocity panel -111.6 \kms marks two long structures that resemble intergalactic v-shaped filaments and an \hi\ knot is identified in panel -106.7 \kms\ that may be emission associated with the filamentary structures.  
Velocity panels -110.8 \kms\ and -105.9 \kms\ identify \ion{H}{1} emission that is likely foreground and associated with the Milky Way warp.  The \hdeg\ and filaments and their association with NGC 1569 (or lack thereof) is discussed in detail in Section \ref{sec:2.1}.
 
 Figure \ref{fig:2} shows \ion{H}{1} emission from three single channels that contain Milky Way emission to demonstrate the range of Milky Way structures and intensities seen throughout the data.
These three channels are at velocities between NGC 1569/UGCA 92 and IC 342/UGCA 86.  The locations of the four galaxies are identified by the contours.  Also, the bottom two panels of Figure \ref{fig:2}, show part of the \hi\ disk of IC 342 that is highly obscured by the Milky Way but is visible because of the high velocity resolution of the GBT data.
 
 Figure \ref{fig:3} shows three individual channels that contain IC 342 and UGCA 86 and their respective \hi\ structures.  
An \ion{H}{1} `spur' attached to the north of UGCA 86 as described by \citet{sti05}, is identified in the bottom two panels.  Two newly discovered \hi\ tails associated with IC 342 and UGCA 86, respectively, are marked in the top panel.
The intensity scale in Figure \ref{fig:3} is in \jybm\ and was chosen so that this \ion{H}{1} spur can be observed.  

Figure \ref{fig:u86ic342} shows an integrated intensity map that was created by summing over 153 \kms\ from velocity channels 36.5 to 188.8 \kms.  This map displays the newly discovered tidal tails associated with UGCA 86 and IC 342, which may form a continuous \hi\ bridge that extends off the southern edge of the map.  The implications of these new features are discussed further in Section \ref{sec:2.2}.  

There is an unidentified \hi\ cloud near (145$\fdg7$, 9$\fdg9$) marked in Figure \ref{fig:u86ic342}.  This cloud has a peak intensity of 0.6 K, a velocity width of $\sim$10 \kms\ (from 33.3 to 43.0 \kms), and an integrated column density of $N_{\rm HI}$ = 2.9 $\times$ 10$^{18}$ cm$^{-2}$.  The origins of this cloud are currently unknown.  It may be foreground emission from the Milky Way, although it is offset by 11 \kms\ in velocity from the Milky Way, or it could be a dwarf galaxy, although there is no known stellar component (STScI Digitized Sky Survey (DSS), 
The Two Micron All Sky Survey (2MASS), Wide-field Infrared Survey Explorer (WISE), and the NASA Extragalactic Database (NED)
were searched; this region is outside of the Sloan Digitized Sky Survey (SDSS) footprint).  If this cloud were at the average distance of the IC 342 galaxy group, 3.28 Mpc \citep{kar03}, then it would contain an \hi\ mass of 1.3 $\times$ 10$^7$ $M_{\sun}$. No distance information can be obtained from the data in hand and further studies are needed to determine the nature and origin of this cloud.  Therefore, the mass estimate is not included in Table \ref{tab:2}.

Table \ref{tab:2} lists average column densities and total \ion{H}{1} masses for all the features described. The AIPS task {\sc ispec} was used to integrate rectangular regions of the map over consecutive channels in order to calculate the total flux of the features.  These fluxes were then converted into average \ion{H}{1} column densities and masses, assuming the distances in Table \ref{tab:1}. The spiral galaxy IC 342 does not have a column density or \ion{H}{1} mass stated because a large part of IC 342 is masked by the Milky Way.

\section{Analysis and Discussion}\label{sec:2}

The purpose of these GBT observations is to determine if there is extended \ion{H}{1} emission connecting NGC 1569 to members of the IC 342 galaxy group.  IC 342 is approximately 290 kpc at a distance of 3 Mpc, projected on the sky, from NGC 1569. 
 At the sensitivity limit of $N_{\rm HI}$ = 1.4 x $10^{18}$ cm$^{-2}$ (5$\sigma$ for 20 \kms\ line width), no extended \ion{H}{1} structures connecting IC 342 or UGCA 86 to NGC 1569 were detected.  However, Milky Way emission lies between NGC 1569 and IC 342/UGCA 86 in velocity, 
so there could be tenuous structures connecting these objects that are blended with Milky Way \hi\ emission. The data are analyzed in two parts, first NGC 1569 and UGCA 92 are discussed, and second, IC 342 and UGCA 86 are examined.
 
 	\subsection{NGC 1569 and UGCA 92}\label{sec:2.1}
 
 		\subsubsection{Extended \ion{H}{1} features}\label{sec:2.1.1}

Two distinct structures were detected that are suggestive as having a connection to NGC 1569.  The first is a large \ion{H}{1} cloud with a diameter of 0$\fdg$5 (about 26 kpc) that has the same position and velocity as NGC 1569.  The cloud is elongated and extends south toward UGCA 92.  It is visible in $\sim$6 channels in Figure \ref{fig:1}, giving it a velocity width of only $\sim$4 km s$^{-1}$ and it has an average column density of $N_{\rm HI}$ = 5.7 x 10$^{18}$ cm$^{-1}$.

Figure \ref{fig:4} shows velocity channel -113.17 km s$^{-1}$, which displays the prominent emission of the 0$\fdg$5 \ion{H}{1} cloud.  The white contours in the left image of Figure \ref{fig:4} are high resolution VLA integrated \ion{H}{1} flux contours of NGC 1569 (Paper I) that are over plotted on the GBT map for identification and to show the morphology.  The right image shows the integrated intensity map (Figure 5, Paper I).  The VLA was able to resolve the strongest emission of the \hdeg, as seen by the diffuse teardrop-shaped emission extending to the south of the galaxy disk.  Also prominent in the VLA intensity map is tenuous emission stretching northeast of the disk, which is suggestive of a counter \hi\ ``tail''. The intensity scales in Figure \ref{fig:4} are shown in M$_{\sun}$ pc$^{-2}$ for comparison and they are drastically different because the GBT image on the left is only a single 0.81 \kms\ channel while the VLA image on the right is the fully integrated intensity field.

The second features discovered are two filaments joined together in a v-shape extending out from NGC 1569 and the \hdeg\ in the direction of UGCA 92.  These filaments are about 1$\fdg$5 long stretching nearly the entire projected distance to UGCA 92, which is roughly 77 kpc in length projected on the sky at the distance of NGC 1569.  The filaments have an average \hi\ column density of 4.9 x $10^{18}$ cm$^{-2}$ and an integrated mass of 5.6 x $10^7 M_{\odot}$.  Figure \ref{fig:bil} shows velocity channel -110.8 \kms\ where the filaments are obvious.  These filaments appear at more positive (red-shifted) velocities than the 0$\fdg$5 cloud.  An angle of $\sim27\arcdeg$ projected on the sky separates the two filaments.  They are only visible in 6 channels and, therefore, have the same velocity width of $\sim$4 km s$^{-1}$ as the 0$\fdg$5 \ion{H}{1} cloud. There is a knot of \ion{H}{1} emission that lingers at more red-shifted velocities than the filaments (in velocity channels -107.5 to -105.1 \kms), as identified in Figure \ref{fig:1}.  It is not clear if this emission is part of the filaments or not.  Table \ref{tab:2} shows the average column density and mass of the filaments, where the knot of \ion{H}{1} emission in velocity panels -107.5 to -105.1 \kms\ was included.  

From the GBT data presented here, a total \ion{H}{1} mass for NGC 1569 of 2.4 x $10^8 M_{\odot}$ was derived, assuming a distance of 2.96 Mpc. This \ion{H}{1} mass agrees with the \ion{H}{1} mass of 1.8 x 10$^8 M_\sun$ determined from VLA data from Paper I and the \ion{H}{1} mass of 1.6 x 10$^8 M_\sun$ derived by \citet{muh05} as well as the \ion{H}{1} mass of 2.3 x 10$^8 M_\sun$ determined by \citet{sti02} (masses corrected for the adopted distance).  For the 0$\fdg$5 \ion{H}{1} cloud, an average column density of 5.7 x $10^{18}$ cm$^{-2}$ and a mass of 2.6 x $10^7 M_{\odot}$ were calculated.  The 0$\fdg$5 \ion{H}{1} cloud has a mass about 10\% of the total \ion{H}{1} mass of NGC 1569.   
A total combined mass for both the 0$\fdg5$ \ion{H}{1} cloud and the filaments is 8.2 x $10^7 M_{\odot}$,  
roughly 30\% of the \ion{H}{1} mass of NGC 1569. 

These GBT data produce an \hi\ mass of 2.0 $\times$ 10$^{8}$ $M_\odot$ for UGCA 92, compared to 1.56 $\times$ 10$^{8}$ $M_\odot$ found by \citet{beg08a}.  This discrepancy may be due to foreground Milky Way emission that was added into the GBT mass.  It is not possible to separate foreground Milky Way from UGCA 92 or NGC 1569.

  		\subsubsection{Milky Way versus Extragalactic \ion{H}{1} Emission}\label{sec:2.1.2}

		
The top panel of Figure \ref{fig:6} shows a position-velocity diagram sliced through the data cube along the black line shown in the bottom panel, which is an integrated intensity map that includes NGC 1569 and UGCA 92.  The 0$\arcmin$ angular offset on the x-axis of the top panel coincides with the center of the black line indicated in the bottom panel.  The 0$\fdg$5 \ion{H}{1} cloud is marked in the top panel as well as the Milky Way cloud, shown in Figure \ref{fig:4}, that resides just south of UGCA 92.  The horizontal black line in the top panel of Figure \ref{fig:6} is to guide the eye along the -113 \kms\ velocity line, which marks the velocity channel where the \hdeg\ is strongest in emission.  The Milky Way cloud south of UGCA 92 is red-shifted compared to the \hdeg. Also marked in the top panel of Figure \ref{fig:6}, is the Milky Way and Milky Way warp velocity ranges.  Both NGC 1569 and UGCA 92 have velocities that overlap with the Milky Way and Milky Way warp, as evidenced by the vertical light yellow features that stretch upwards at the locations of NGC 1569 and UGCA 92 into the Milky Way emission where the velocities become more positive.  However, the Milky Way warp at these velocities and positions is tenuous, as seen by the low flux of the Milky Way warp region marked in Figure \ref{fig:6}, and the Milky Way disk emission is significantly offset in velocity from the 0$\fdg$5 \ion{H}{1} cloud and filamentary features.  

The systemic velocity of UGCA 92 is -99 km s$^{-1}$, which coincides with the upper limit of the Milky Way warp velocity range.  NGC 1569 has a systemic velocity of -85 km s$^{-1}$, which is between the peak Milky Way and Milky Way warp velocities, although it is likely that there is some Milky Way emission at this velocity also.  Since both NGC 1569 and UGCA 92 have velocities in the range of the Milky Way and the Milky Way warp, one would expect any extended \ion{H}{1} features associated with either of these galaxies to also have velocities confused with the Milky Way.  So how can one tell which features are Milky Way and which are associated with NGC 1569 and/or UGCA 92?

One way to approach this problem is to look at the line profiles of different \ion{H}{1} features in the area around NGC 1569 and UGCA 92.  Figure \ref{fig:7} shows five line profiles of a single pixel each through the centers of different \hi\ features near NGC 1569. The plots are
at the velocities of the extended \ion{H}{1} features in question and are plotted on the same temperature intensity scale.  
The exact positions of the line profiles are given in the upper left corner of the plots and correspond to the positions shown in the intensity image in Figure \ref{fig:7}(f).  
The 0$\fdg$5 \ion{H}{1} cloud has a line profile that is narrower and has a lower intensity than the two Milky Way clouds in Figures \ref{fig:7}(b) and \ref{fig:7}(c).  Also, the wings of the line are broader than the Milky Way clouds' line profiles, which could be from a mix of warm and cold \hi\ that one would expect associated with a galaxy such as NGC 1569 \citet{you01}.  Foreground Milky Way clouds along the line of sight contain only one temperature component \citet{for10} and thus, would not have broad wing features.  

The Milky Way cloud in Figure \ref{fig:7}(b), is a Gaussian-like profile with a peak intensity at around 4.5 K, compared to a peak intensity of only 2.7 K for the 0$\fdg$5 \ion{H}{1} cloud in \ref{fig:7}(a).  Figure \ref{fig:7}(c) shows a line profile of a Milky Way cloud located at the bottom of the GBT map below NGC 1569.  This profile shows the cloud peak intensity is 4.2 K and the FWHM is broader compared to the 0$\fdg$5 \ion{H}{1} cloud in \ref{fig:7}(a).  The velocity widths of the line profiles for the three clouds can also be seen in Figure \ref{fig:1}.  The 0$\fdg$5 \ion{H}{1} cloud is observed in only 6 channels ($\Delta V$ $\sim$ 4 km s$^{-1}$) while the Milky Way cloud directly below UGCA 92 and the Milky Way cloud at the bottom of the map are seen in about 16 channels total ($\Delta V$ $\sim$ 12 km s$^{-1}$), including two channels beyond what is shown in Figure \ref{fig:1}.  

In addition, both of the line profiles for the two Milky Way clouds are slightly red-shifted compared to the 0$\fdg$5 \ion{H}{1} cloud.  This shift in velocity is also observed in Figure \ref{fig:1} and Figure \ref{fig:6}. Both Milky Way clouds have the same velocity range, they both appear in velocity channel -114.0 \kms\ and disappear in velocity channel -100.3 \kms, which is beyond Figure \ref{fig:1}.  The 0$\fdg$5 \ion{H}{1} cloud is slightly blue-shifted compared to both of these Milky Way clouds, as Figure \ref{fig:1} shows it first appearing in velocity channel -114.8 \kms, compared to -112.4 \kms\ for the Milky Way clouds.  Although this is only a 2.4 km s$^{-1}$ shift, the peaks of the line profiles for these Milky Way clouds in Figure \ref{fig:7} are offset by about 2.5 km s$^{-1}$ from the 0$\fdg$5 \ion{H}{1} cloud, in agreement with Figure \ref{fig:1}. 

The centers of NGC 1569 and UGCA 92 are also plotted in Figures \ref{fig:7}(d) and \ref{fig:7}(e), respectively.  These line profiles are broad, as expected for rotating galaxies, but NGC 1569 shows a very narrow component at the peak of the flux.  This narrow peak corresponds with the peak velocity of the 0$\fdg$5 \ion{H}{1} cloud.  It appears that the 0$\fdg$5 \ion{H}{1} cloud overlaps with NGC 1569 in both position and velocity.  In Paper I, the VLA observations show high velocity \ion{H}{1} surrounding the galaxy as well as tenuous emission to the south and north (see Figure \ref{fig:4}).  All of this \ion{H}{1} emission detected by the VLA, matches the velocities of the 0$\fdg$5 \ion{H}{1} cloud in our GBT map.  Perhaps the 0$\fdg$5 \ion{H}{1} cloud is the source for these high velocity \ion{H}{1} features seen surrounding NGC 1569 in the VLA data.

The filaments need a different analysis, as they are elongated features without a central core of concentrated \ion{H}{1} emission.  Beginning with Figure \ref{fig:1}, the filaments are observed in only about seven channels, similar to the 0$\fdg$5 \ion{H}{1} cloud and they extend out of this cloud both spatially and in velocity forming a continuous velocity bridge connecting the disk of NGC 1569, \hdeg, and ending with the filaments.  Also, there are no other v-shaped features like these anywhere else in the data.  Tenuous wisps are seen throughout Figure \ref{fig:2} but, none match the distinctive v-shape that the filaments exhibit.  Furthermore, the other wisps and filament-like structures seen in Figure \ref{fig:2} are observed in more than just six channels and are much longer in length.  
Figure \ref{fig:bil} displays the entire mapped region for the velocity channel that contains the maximum emission from the filaments.  The Milky Way cloud directly south of UGCA 92 is prevalent and there are other tenuous filamentary structures visible in this field.  Even though the v-shape is distinct, it is not possible to directly separate foreground Milky Way emission from extragalactic emission at the velocities of the filaments or \hdeg.

As for other structures that may be contaminating the field such as high velocity clouds (HVCs) around the Milky Way, the only likely candidate is HVC Complex H.  However, according to an \ion{H}{1} map of HVC Complex H by \citet{wak98}, the HVC does not overlap with NGC 1569 or UGCA 92 in position or velocity. There are no other HVCs in the vicinity of NGC 1569 that match position or velocity. 

Although it is impossible to say definitively that the 0$\fdg$5 \ion{H}{1} cloud and filaments are not part of the Milky Way, Milky Way warp, or any HVC associated with the Milky Way, the evidence (blue-shifted velocity, narrow width, and wide wings of the line-profile of the 0$\fdg$5 \ion{H}{1} cloud compared with other nearby Milky Way clouds, and the distinctive v-shape of filaments) suggests that these structures are extragalactic in nature.  

Another argument can be made in support of the extragalactic nature of the \hdeg\ and filaments from \citep{gro12}.  They detect an older stellar population in NGC 1569 that stretches 6$\arcmin$ in the same direction and position of the \hdeg.  This stellar halo is suggestive of a tidally disturbed stellar disk, of which the implications will be discussed in Section \ref{sec:3}.  

	\subsection{IC 342 and UGCA 86}\label{sec:2.2}
	
As previously stated, Milky Way \ion{H}{1} emission lies between NGC 1569/UGCA 92 and IC 342/UGCA 86 in velocity and, in fact, nearly half of IC 342's rotating \ion{H}{1} disk is confused with Milky Way velocities.  Figure \ref{fig:8} shows a position-velocity diagram, similar to the one in Figure \ref{fig:6}, that slices through the centers of UGCA 86 and IC 342.  The emission feature through IC 342 is extended in position and velocity due to the rotation of the disk.  

In addition to the position-velocity diagram, line profiles were made through the centers of UGCA 86 and IC 342, as shown in Figure \ref{fig:9}.  These profiles show the extent of the Milky Way at velocities below 0 km s$^{-1}$ and show that IC 342 and UGCA 86 are well separated from Milky Way confusion above 0 km s$^{-1}$.  Also, there are no HVCs known that match the positions and velocities of IC 342 and UGCA 86 \citep[see e.g.,][]{bra99, loc02, loc03, sim06}. 
Therefore, it is assumed that all \ion{H}{1} emission above 0 km s$^{-1}$ is associated with IC 342 and UGCA 86.  

From these GBT data, an \ion{H}{1} tail was detected extending from the south to the west from UGCA 86.  \citet{sti05} identified an \ion{H}{1} spur to the north of the galaxy in their high resolution \ion{H}{1} data taken with the Dominion Radio Astrophysical Observatory (DRAO) Synthesis Telescope.  They argued that the spur came from an interaction with IC 342, although this was difficult to prove partly because no counter tail was observed in their data.  Figure \ref{fig:3} shows individual channel maps at low signal-to-noise for the purpose of displaying the counter tail to the \ion{H}{1} spur in UGCA 86.  This counter \ion{H}{1} tail is very tenuous with an average column density of only 4.2 x $10^{18}$ cm$^{-2}$ and a mass of 2.6 x 10$^7 M_\sun$.  The \ion{H}{1} tail in UGCA 86 is observed over a velocity range of $\sim$30 to $\sim$78 km s$^{-1}$.  This \ion{H}{1} tail is about 36$\arcmin$ long, which at the distance of UGCA 86, is about 31 kpc projected on the sky.  The \ion{H}{1} spur feature, identified in Figure \ref{fig:3}, is observed over a velocity range of $\sim$66 to $\sim$156 km s$^{-1}$.  The spur is about 16$\arcmin$ in length, which is $\sim$14 kpc projected on the sky. It seems plausible that these features are related because of the continuity in the velocities between them.  The systemic velocity of UGCA 86 is 72 km s$^{-1}$ (see Table \ref{tab:1}), and this lies between the velocities of the two tail-like features.  

Table \ref{tab:2} lists the mass of UGCA 86 that was determined by summing over pixels that contained the body of the galaxy disk and did not include pixels that contained emission from the spur or tail.  Also, no channels below 10 \kms\ were included in the calculation because of strong Milky Way emission below this velocity.  The GBT data give a mass of 4.8 $\times$ 10$^8$ $M_\odot$ for UGCA 86, which is comparable to the mass found by \citet{sti05} of 4.4 $\times$ 10$^8$ $M_\odot$, corrected for distance and subtracted for the spur contribution.  \citet{sti05} give a total dynamical mass of 2.3 $\times$ 10$^{10}$ $M_\odot$, again corrected for distance. 

IC 342 also shows a surprising \ion{H}{1} feature extending to the south.  The tail-like structure connected to IC 342 seen Figure \ref{fig:3} is very tenuous but persists in roughly 65 channels over a velocity range of $\sim$32 to 84 km s$^{-1}$, which matches the velocity range of the newly discovered tail south and west of UGCA 86.  A column density of 3.3 x $10^{18}$ cm$^{-2}$ and a mass of 1.6 x 10$^7 M_\sun$ was calculated for this \ion{H}{1} tail and it extends about 0$\fdg$75 on the sky for a physical distance of approximately 40 kpc, at the distance of IC 342.  Figure \ref{fig:u86ic342} displays the integrated \hi\ map highlighting the tails extending from UGCA 86 and IC 342.  

		\subsubsection{Could the two tail-like features attached to UGCA 86 and IC 342 be related?}\label{sec:2.2.1}
		
It has been previously suggested that IC 342 and UGCA 86 are interacting \citep{rot79,cro00,gra02,sti05}, and if the two tails are related, then this would provide concrete evidence that IC 342 and UGCA 86 are indeed an interacting pair of galaxies.  
Although the data do not reveal an \ion{H}{1} bridge connecting the two tails of UGCA 86 and IC 342 at the sensitivity limit, it is possible that these features form a continuous \ion{H}{1} stream because they share the same velocities and relative position to one another.  The connecting \hi\ bridge may lie off the edge of the map to the south of the two galaxies.  Figure \ref{fig:u86ic342} shows an integrated intensity map highlighting the two tails and tenuous emission can be seen extending off the edge of the map below IC 342. 

If these two tails do not comprise a continuous \ion{H}{1} stream, then they could either be old tidal features in the process of dissipating and falling back onto their host galaxies \citep[][and references therein]{hib00} or they could be relatively new features still in the process of forming a continuous bridge \citep{ton10}.  

According to models of interacting galaxies made by \citet{bar09}, short, compact features indicate a recent interaction while extended tails having the same velocities could mean either the objects are near apocenter or the velocity vectors of the tidal tails are perpendicular to the line of sight.  UGCA 86 has a short, pronounced \ion{H}{1} spur to the north of the disk, as discussed previously and shown in Figure \ref{fig:3}, which gives credence to the interaction being recent.  Also, UGCA 86 is experiencing an episode of enhanced star formation but not a global starburst \citep{sti05}.  \citet{hib00} explains that the timescales for interactions and subsequent starburst activity are quite different, as observed in systems like Arp 299 and the Antennae.  The tidal tails in these large galaxies stretch for hundreds of kpcs and formed hundreds of millions of years ago while the starburst activity occurred only tens of millions of years ago and may be episodic \citep{hib00}.  Because of the recurring nature of the starbursts, one cannot use the presence or absence of a starburst at the current time as an indication of when an interaction occurred.  

Staying with the assumption that there is no connecting \hi\ bridge lying off the edge of the map, the tidal tails of UGCA 86 and IC 342 are only a projected length of 31 and 40 kpc, respectively, which may mean that they are young features that haven't had enough time to extend fully \citep{ton10}.  If they were old features in the process of dissipating and falling back onto their host galaxies, then one would expect to see remnants of fine structure such as shells or loops \citep{hib00}, assuming that the potentials in both galaxies are relaxed.  A rough estimation for the dynamical timescale of an interaction between UGCA 86 and IC 342 can be made by using the velocity dispersion of the galaxy group, which is $\sim$ 54 \kms\ \citep{kar05}.  Using the projected distance of $\sim$87 kpc, UGCA 86 passed through pericenter $\sim$1.5 Gyr ago.  Mapping the region below IC 342 and UGCA 86 at a lower sensitivity to \hi\ column density can be useful in determining the interaction history for these two galaxies.

\section{Exploring the Implications of the Extended \ion{H}{1} Emission Around NGC 1569}\label{sec:3}

NGC 1569 is a complex galaxy with complex stellar and gas kinematics. The stars increase in velocity dispersion in the region of the SSCs compared to the center of the galaxy, and the stellar mass dominates the inner disk's gravitational potential.  The stars and gas are also kinematically coupled.  The \ion{H}{1} data show strong non-circular motions across all of the disk, as detailed from VLA observations from Paper I.  There is a strong non-circular motion \ion{H}{1} cloud to the south of the SSCs that appears to be impacting the galaxy and could be responsible for the starburst and dense \ion{H}{1} ridge and cloud that resides west of the SSCs.  Also, there is tenuous \ion{H}{1} emission detected with the VLA south and north of NGC 1569, and the velocities of this emission match those of the 0$\fdg$5 \ion{H}{1} cloud discovered with the GBT.  The GBT data also show v-shaped filaments extending a projected distance of 88 kpc in the direction of another dwarf UGCA 92. 

These observations paint a picture of a very disturbed dIm galaxy and prompt three possible explanations for the evolution that lead to the system seen today:\\
1) The extended \ion{H}{1} features observed in the GBT map are primordial gas structures from cosmological filaments that are accreting onto NGC 1569.  This gas falls into the inner disk thereby creating the disturbed kinematics and starburst.\\
2) NGC 1569 is a merger remnant of two gas rich dwarf galaxies.  \\
3) NGC 1569 is interacting with UGCA 92 during which, the quiescent outer \ion{H}{1} disk of NGC 1569 was disrupted and caused this gas to lose angular momentum and fall to the center.

The possibility that the filaments and 0$\fdg$5 \ion{H}{1} cloud seen in the GBT data are primordial gas from cosmological filaments is unlikely. In the early universe, primordial gas accretes onto galaxies through large-scale filamentary structures, which may be present around NGC 1569.  However, the chances of this neutral \hi\ gas surviving in a dense galaxy group to the present time is highly improbable.  Intergalactic UV photoionization causes low column density \ion{H}{1} gas to become ionized and evaporate over a relatively short timescale \citep[see e.g.,][]{bor10, hib00, mal93, san08}.  The average column density of the \ion{H}{1} filaments extending from NGC 1569 is 4.9 x 10$^{18}$ $M_\sun$.  Therefore, a primordial origin for these structures from cosmological filaments is disregarded.

A merger is more plausible since merger remnants of large disk galaxies have similar characteristics as NGC 1569, particularly the filaments and starburst activity, only on larger scales.  
For example, the ongoing merger Arp 299 shows very long, parallel \ion{H}{1} tidal tails that share the same velocities \citep{hib99}, which are similar to the filaments detected in NGC 1569.  Arp 299 is also undergoing an intense starburst, and has very disturbed kinematics in both stars and \ion{H}{1} \citep{hib99}.   The mass ratio between the \hi\ disk of Arp 299 and the \hi\ tails is $\sim$2.6, using the disk mass corrected for absorption from \citet{hib99}.  NGC 1569 has a ratio of \hi\ mass in the disk to \hi\ mass of the \hdeg\ plus v-shaped filaments of $\sim$2.9, which is comparable to Arp 299.  However, there are major differences between Arp 299 and NGC 1569.  Arp 299 has {\it stellar} tidal tails and also contains two distinct nuclei \citep{hib99}.  \citet{gro12} find a stellar halo of old stars that reaches 6$\arcmin$ south in the same direction as the \hdeg\ cloud.  This is suggestive as being a tidally disturbed stellar disk, but there have been no studies to date designed to look for a low brightness stellar population in the filaments of NGC 1569 and a double nucleus is not observed.  

Another example of a large disk merger remnant with two tidal filaments is NGC 1052.  There are two \ion{H}{1} filaments extending out from NGC 1052 \citep{van86} and it is also undergoing a starburst \citep{pie05}. However, NGC 1052 has an ionized gas disk that rotates perpendicular to the stellar disk \citep{van86}, and has an E4 morphology (NED).  One of the striking results from Paper I is that the stars and the gas of NGC 1569 kinematically follow each other, which is not what is observed in NGC 1052.  NGC 1569 does not have an elliptical shape but rather, is a thick disk. NGC 1052 also has an active nucleus and NGC 1569 does not.

The merging scenario to explain the observed features of NGC 1569 is a compelling theory, but still there are obvious drawbacks when compared to larger merger remnant galaxies.  In the case of Arp 299, this merging system shows two distinct nuclei so, perhaps it is at an earlier stage of merging than NGC 1569.
On the other hand, dwarf galaxies don't have nuclei so comparing these two objects may be crude at best.  Large spiral disk galaxies are more complex with supermassive black holes in their centers, spiral arms, and bulges in some that all combine for a more dramatic merging event than is expected in the dwarf galaxy regime.  Gas rich dwarfs do not have the sophisticated structure of their spiral galaxy counterparts nor do they contain supermassive black holes that can ignite and become active during a merging event as is the case of NGC 1052.  But no one, as yet, has modeled dwarf-dwarf galaxy mergers in isolation so understanding the morphological changes that can occur in these systems is unknown.  Spiral-spiral galaxy mergers generally produce elliptical systems.  Could this be the case for two merging disk dwarf galaxies, too?  If so, then perhaps the E4 galaxy NGC 1052 is at a later stage of merging than NGC 1569, which may be on its way to becoming a dE (dwarf elliptical) system.

There may be other evidence that NGC 1569 is a merger remnant from a study by \citet{sti98} where they describe a potential \ion{H}{1} companion.  
Is it possible that this \ion{H}{1} companion is a gas remnant from a dwarf-dwarf galaxy merger that has formed NGC 1569?  Stil \& Israel suggest that the \ion{H}{1} companion is either primordial gas settling into the potential well of NGC 1569 or it is a separate merging object.  Cosmological filaments carrying primordial gas as the source for the 0$\fdg$5 \ion{H}{1} cloud and v-shaped filaments seen in the GBT data has already been disregarded, so it does not seem likely that the possible \ion{H}{1} companion discussed by Stil \& Israel is primordial in this sense. However, the \ion{H}{1} companion could be part of a merging event that has created the starburst in NGC 1569.  The velocities of this \ion{H}{1} companion match the velocities of the GBT 0$\fdg$5 \ion{H}{1} cloud, which suggests that these structures may have a similar origin.  If so, then it is also possible that the 0$\fdg$5 \ion{H}{1} cloud came from a recent interaction with UGCA 92, thus, the \ion{H}{1} companion may have formed from an interaction rather than a merger.
Since no assumptions can be made about the initial conditions of the progenitors that may have merged to form NGC 1569, one can only theorize where this possible \ion{H}{1} companion originated.  

The observations also suggest that NGC 1569 is interacting with UGCA 92.
\citet{beg08a} find an \hi\ diameter for UGCA 92 of 9$\arcmin$ and a rotation velocity, corrected for internal pressure, of 37.08 \kms.  Thus, a total dynamical mass for UGCA 92 is 1.2 $\times$ 10$^9$ $M_\odot$.  Paper I finds a dynamical mass of NGC 1569 of 1.0 $\times$ 10$^9$ $M_\odot$ (corrected for distance), nearly equal to UGCA 92.  However, when compared to the baryonic mass, UGCA 92 has a baryonic-to-dynamic mass ratio, $M_{\rm Bar}$/$M_{\rm Dyn}$ = 0.18 while NGC 1569 has $M_{\rm Bar}$/$M_{\rm Dyn}$ = 0.37.  This means that NGC 1569 has roughly twice the amount of baryons for approximately the same amount of dynamical mass.  Could UGCA 92 have lost some of its baryons to NGC 1569 during an interaction?  A more detailed analysis of UGCA 92 is required in order to determine if this is the case.

An example of an interacting large spiral galaxy is NGC 4631.  It is an intriguing analogue for comparison to NGC 1569 because it also has two \ion{H}{1} filaments joined in a v-shape which extend out toward a nearby companion, NGC 4656 \citep[see Figure 1 in][]{wel78}.
NGC 4631 is also part of a galaxy group and is experiencing a starburst. The ratio of \hi\ mass in the disk of NGC 4631 to the total \hi\ mass of the filamentary features in the outskirts of the galaxy is $\sim$5, according to \citet{wel78}, and NGC 1569 has an \hi\ disk mass-to-extended feature mass of 2.9.  The similarity in morphology is striking.

According to simulations by \citet{duc08}, the speed of an encounter can affect the morphology of the tidal tails produced.  High-velocity collisions produce tenuous gas tails with non-uniform density.  Duc \& Bournaud give a disk mass-to-tail mass ratio of $\sim$32, which is more than an order of magnitude higher than the 2.9 disk-to-extended \hi\ mass of NGC 1569.  Also, Duc \& Bournaud find that the tidal tails resulting from high-velocity collisions have high velocity gradients, which is completely the opposite for what is observed in NGC 1569.  Therefore, it is unlikely that a high-velocity collision was responsible for the formation of the extended \hi\ found near NGC 1569.

\citet{bar09} state that from their models long and narrow tidal features associated with an interaction indicate an older passage between two interacting systems. Similarly, \citet{hib00} explains that the timescales for a starburst and an interaction can be very different with an interaction taking place hundreds of millions of years in the past while starbursts are episodic with the most recent occurrence taking place only tens of millions of years ago.  In the scheme presented by Barnes \& Hibbard, the long, extended filaments of NGC 1569 are consistent with an older passage by UGCA 92.

The timescale of the most recent starburst episode in NGC 1569 is on the order of 10 Myr and this dIm galaxy has had multiple bursts of intense star formation \citep{ang05}, consistent with the \citet{hib00} models.  
If NGC 1569 and UGCA 92 are interacting, then one can make a rough estimate of how long ago the pericenter approach was by using the GBT projected distance between the objects of $\sim2\arcdeg$, which, at a distance of 2.96 Mpc, corresponds to a physical distance of $\sim$103 kpc.  If NGC 1569 and UGCA 92 are moving through space at approximately the velocity dispersion of the galaxy group, which is 54 \kms\ \citep{kar05}, then it would take about 950 Myr for NGC 1569 to reach its present distance from UGCA 92, assuming NGC 1569 and UGCA 92 are in the plane of the sky.  The timescale of the most recent starburst and the time in which the last interaction could have occurred are roughly consistent with \citet{hib00}. Thus, a picture of NGC 1569 that is roughly consistent with other observations and models of interacting disk galaxies is observed.



\section{Summary \& Conclusions}\label{sec:4}

The GBT data presented here have several exciting implications. Newly discovered \hdeg\ and v-shaped filamentary features were found at the same position as NGC 1569 and contain velocities that form a continuous velocity sequence with the systemic velocity of NGC 1569 having the most red-shifted velocity (-85 \kms), the filaments containing intermediate velocities and the \hdeg\ having the most blue-shifted velocity.  
An analysis of the line profiles of the \hi\ clouds in the field around NGC 1569, display characteristics that are suggestive as being slightly red-shifted from foreground Milky Way and also have narrower peaks and wider wings than Milky Way emission, suggesting a mix of warm and cold \hi\ medium, which is what one would expect in an extragalactic source.  If the \hdeg\ and filaments are, in fact, extragalactic then they are most likely the result of an interaction between NGC 1569 and UGCA 92.  This interaction scenario supports the hypothesis that the triggering mechanism for the recent starburst activity in NGC 1569 is from tidal forces, which created a loss of angular momentum in the outer \hi\ disk, causing material to fall to the center, providing the compression and subsequent cooling of gas necessary for a global starburst (see Paper I for more details).

The results are outlined as follows: \\
1)  
A large, \hdeg, which is 26 kpc in diameter at the distance of NGC 1569, was discovered at the same position and velocity as NGC 1569 and extends to the south toward UGCA 92.  Compared to other foreground Milky Way clouds in the vicinity of NGC 1569, the \hdeg\ has a narrower velocity profile, is slightly blue-shifted from foreground Milky Way emission, and forms a continuous velocity bridge with the \hi\ disk of the galaxy.\\
2)  Two filaments joined together in a v-shape were discovered extending south from NGC 1569 perpendicular to the disk in the direction of UGCA 92.  These filaments have narrow velocity widths and are the most blue-shifted structures that form a continuous velocity sequence  connecting the \hi\ galaxy disk and \hdeg.  \\
3)  The 0$\fdg$5 \ion{H}{1} cloud and filaments are suggestive as extragalactic structures but could be foreground Milky Way.  If they are indeed extragalactic, then they could be evidence for a recent interaction between NGC 1569 and UGCA 92.\\
4)  In the velocity ranges of UGCA 86, a counter tidal tail was discovered on the opposite side of the disk from the \ion{H}{1} spur identified by \citet{sti05}.  This counter tail has velocities that follow a sequential velocity sequence with the spur (which is red-shifted from the systemic velocity of the galaxy), main galaxy body and counter tail (which is blue-shifted from the systemic velocity of the galaxy).  This counter tail is likely associated with the \ion{H}{1} spur, and was probably created in an interaction with IC 342.\\
5) IC 342 also shows a surprising \ion{H}{1} tail-like feature extending south toward UGCA 86.  This \ion{H}{1} tail is very tenuous and at the sensitivity limit of the GBT map, but has a velocity width of over 47 km s$^{-1}$ and has the same velocities as the counter \ion{H}{1} tail discovered in UGCA 86.  The two \ion{H}{1} tails associated with UGCA 86 and IC 342 discovered in the GBT map may form a continuous bridge that extends off the south edge of the map.  More data are needed to confirm this hypothesis.

For future work, I plan to map a deep, $\sim$2$\arcdeg \times 2\arcdeg$ region south of the current area around IC 342 and UGCA 86 and combine these data with the data presented here to test whether the two tidal tails seen in both objects form a tidal bridge.  Also, I intend to investigate what morphologies would be expected from interactions between NGC 1569 and UGCA 92 as well as IC 342 and UGCA 86 using hydrodynamic simulations.  By implementing all necessary physics, including stellar feedback, one can explore the cosmological origin of NGC 1569 and possibly determine its evolutionary fate.

\section{Acknowledgments}

I would like to thank all of the wonderful staff, telescope operators, and astronomers at the Green Bank NRAO site for all of their help and support during the preparation and obtaining of the observations.  In particular, I would like to thank Felix J.\ Lockman for his support, discussions, and advice on this project. I would also like to express my deepest gratitude to Deidre A.\ Hunter for her mentorship, support, and enthusiasm for this work.  Without Deidre, none of this would have been possible.  The LITTLE THINGS team has been a tremendous sponsor of this research and provided many discussions, support and comments, all of which made this paper a success.  I am especially grateful to Elias Brinks, Bruce Elmegreen, Volker Heesen, Andreas Schruba, Kimberly Herrmann, and Hong-Xin Zhang who provided many useful comments and suggestions.
This project was funded by the National Radio Astronomy Observatory's Student Observing Support Award Number GSSP10-0001 and the National Science Foundation under the LITTLE THINGS grant number AST-0707563 to DAH.  This research has made use of the NASA/IPAC Extragalactic Database (NED) which is operated by the Jet Propulsion Laboratory, California Institute of Technology, under contract with the National Aeronautics and Space Administration.

\onecolumn

\begin{figure}
\centering
\includegraphics[scale=0.7]{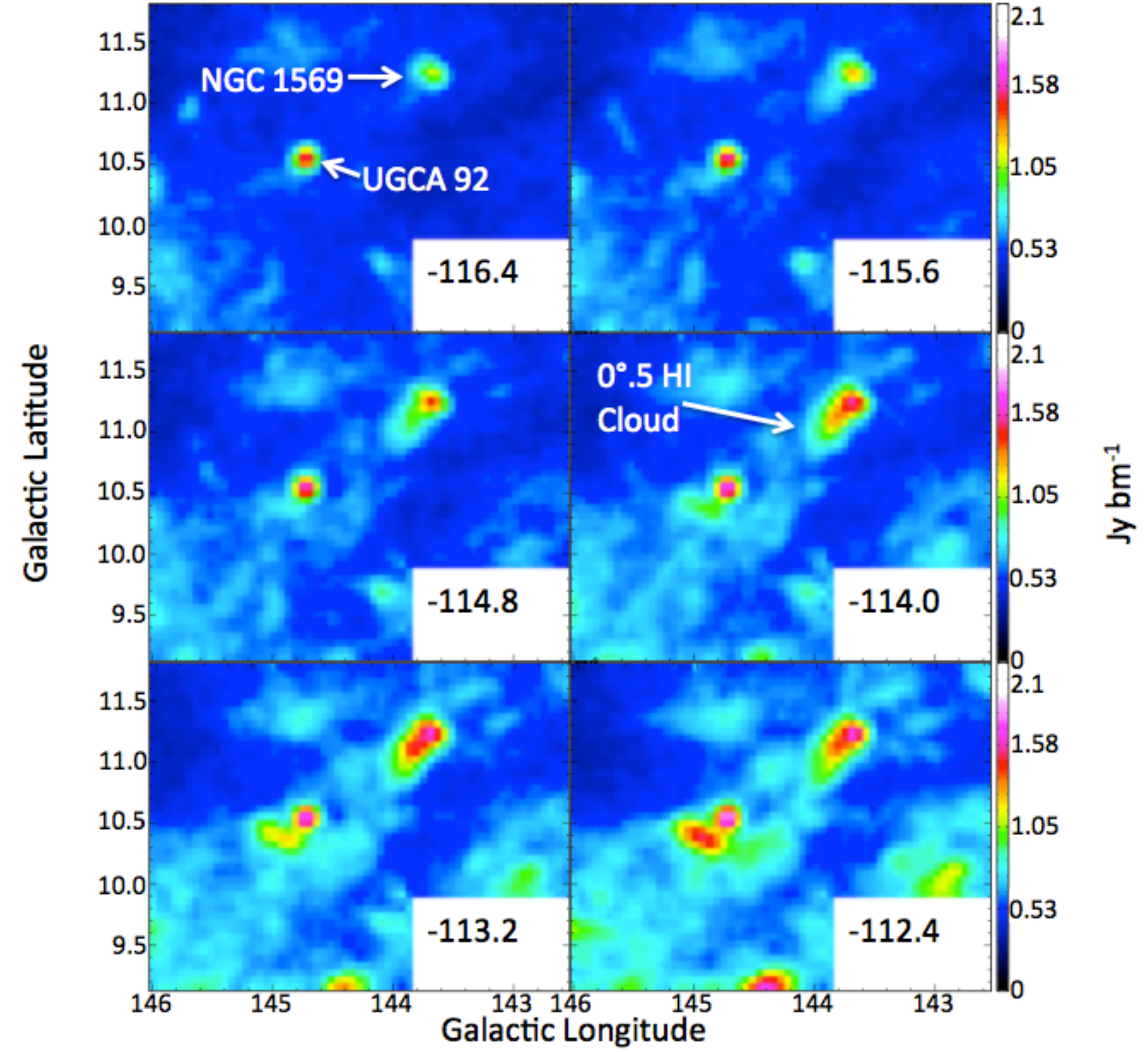}
\caption{Individual channel maps showing NGC 1569 and UGCA 92, identified in the top left panel.  The panels are 0.81 km s$^{-1}$ apart.  The velocity in \kms\ is given in the lower right corner of each panel.  The 0$\fdg$5 \ion{H}{1} cloud is identified in velocity panel -114.0 \kms\ and is visible from -115.6 to -111.6 \kms.  The v-shaped filamentary structures are identified in velocity panel -111.6 and are visible from -112.4 to -107.5 \kms.  There is a knot of \hi\ identified in velocity panel -106.7 \kms\ that may be associated with the filamentary structures and it persists from -107.5 to -104.3 \kms.  
The arrows in velocity panels -110.8 and -105.9 \kms\ show potential Milky Way \ion{H}{1} clouds likely associated with the warp in our Galaxy.  
The intensity bar for each panel is in units of \jybm.}
\label{fig:1}
\end{figure}

\begin{figure}
\centering
\addtocounter{figure}{-1}
\includegraphics[scale=0.7]{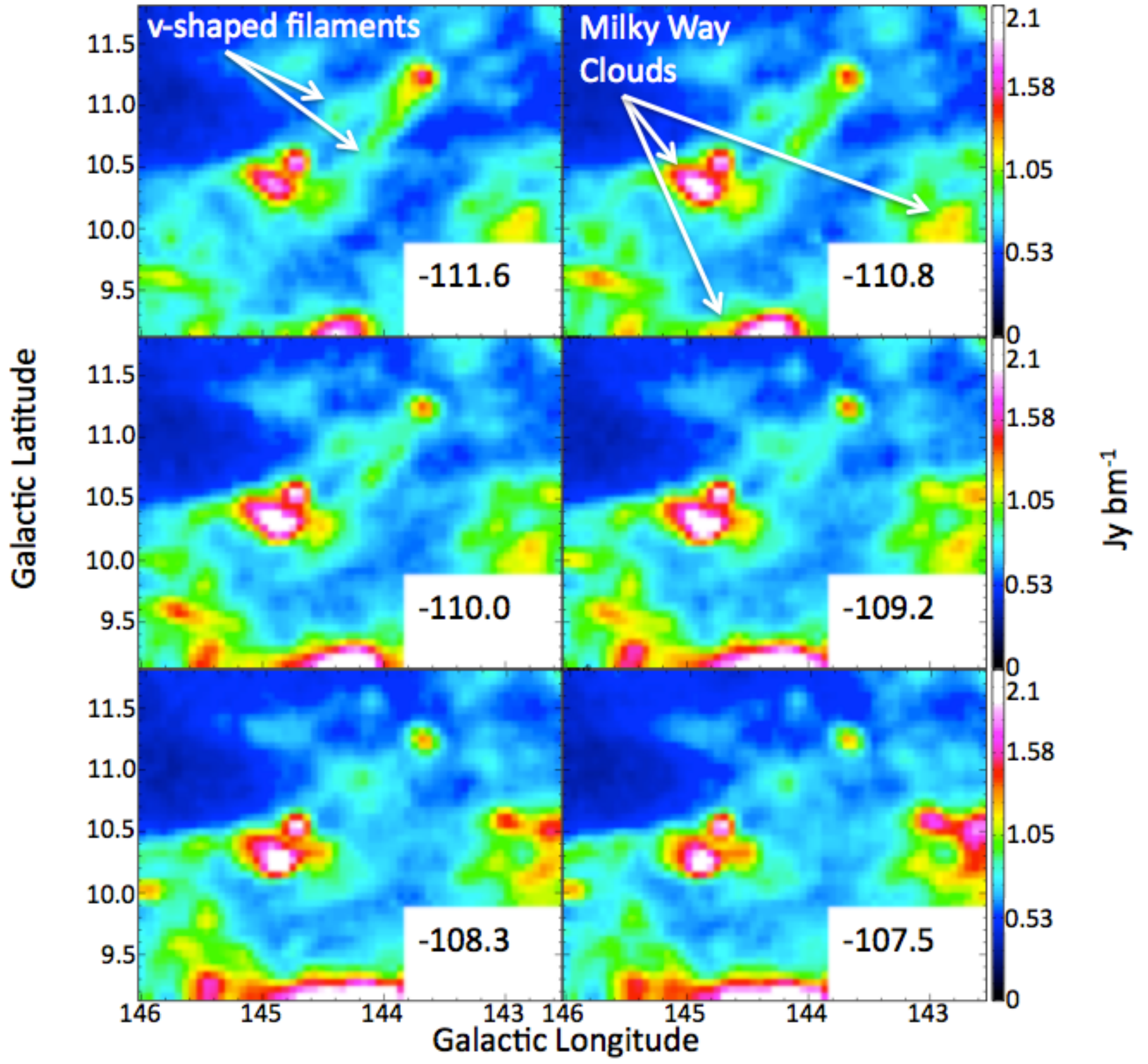}
\caption{Continued}
\end{figure}

\begin{figure}
\centering
\addtocounter{figure}{-1}
\includegraphics[scale=0.7]{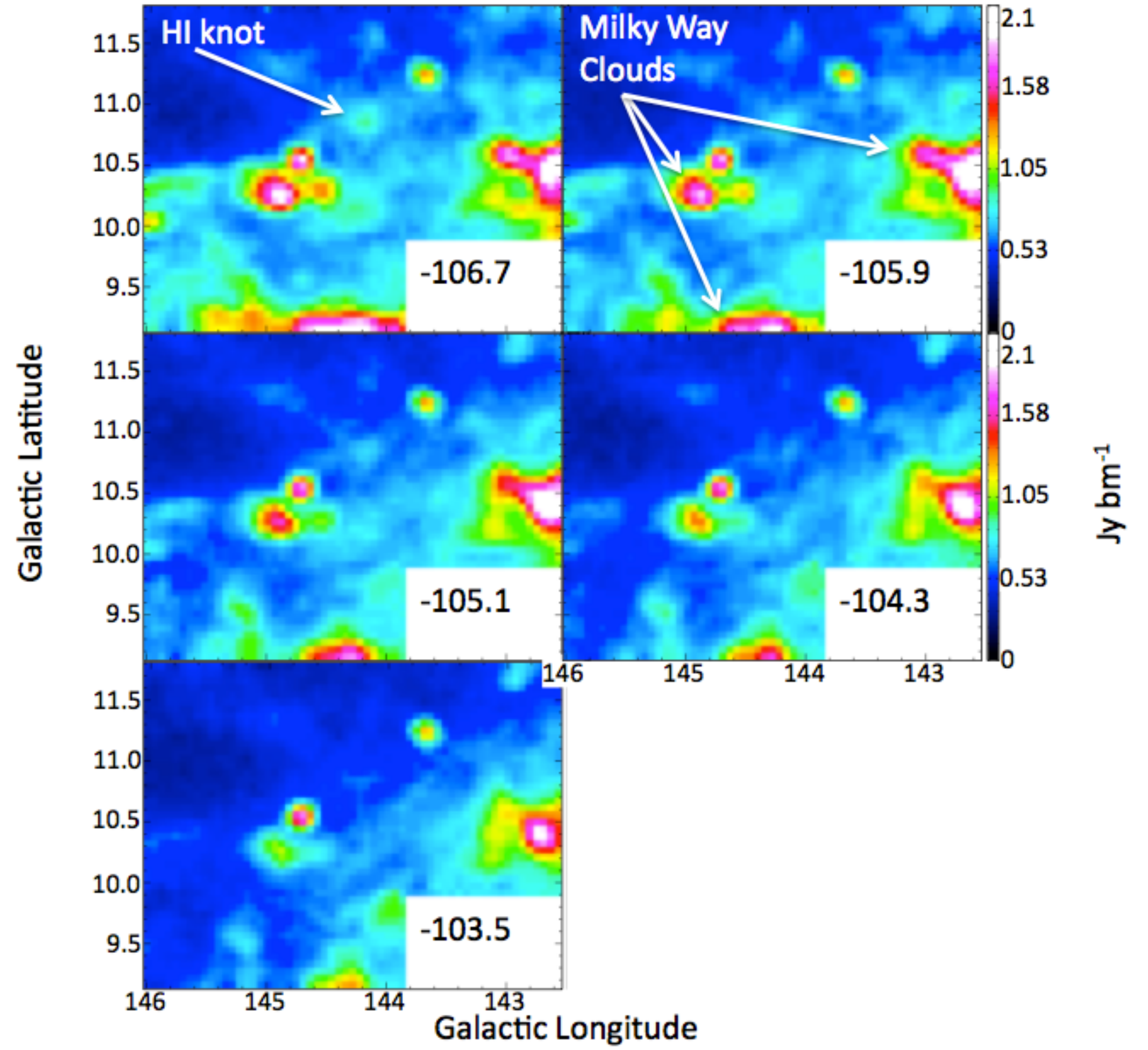}
\caption{Continued}
\end{figure}

\begin{figure}
\centering
\includegraphics[scale=0.4]{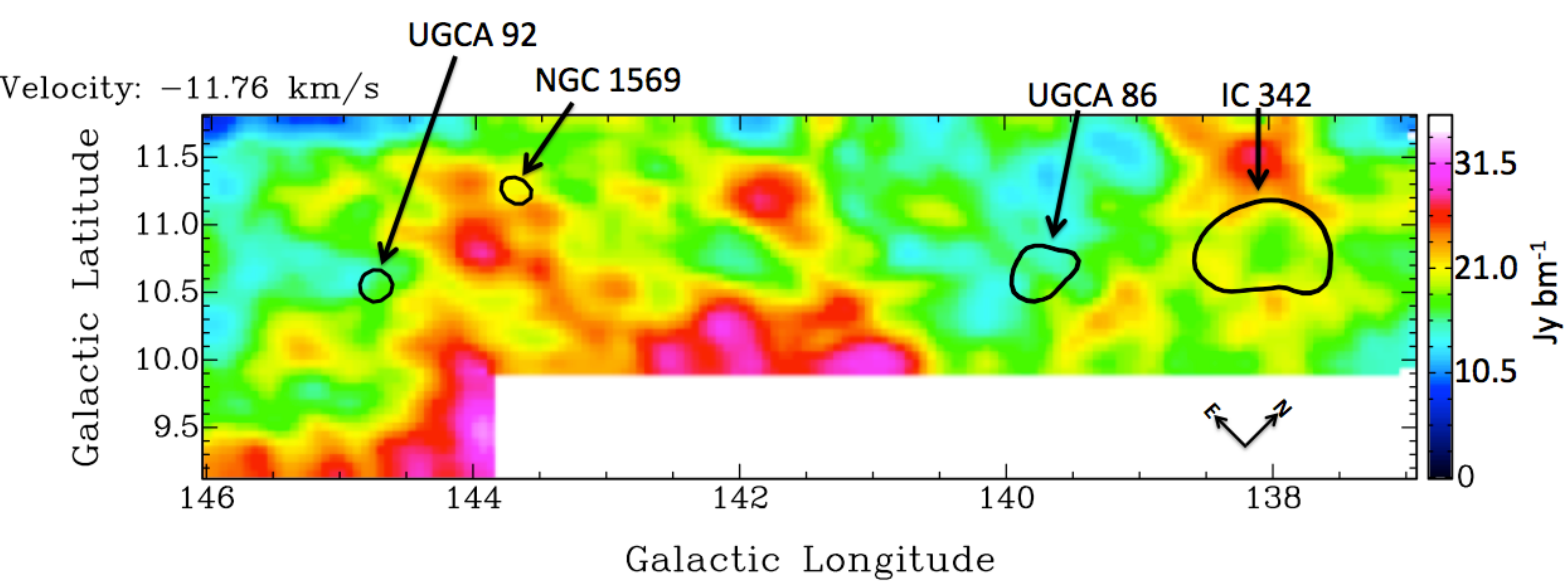}
\hfil
\includegraphics[scale=0.4]{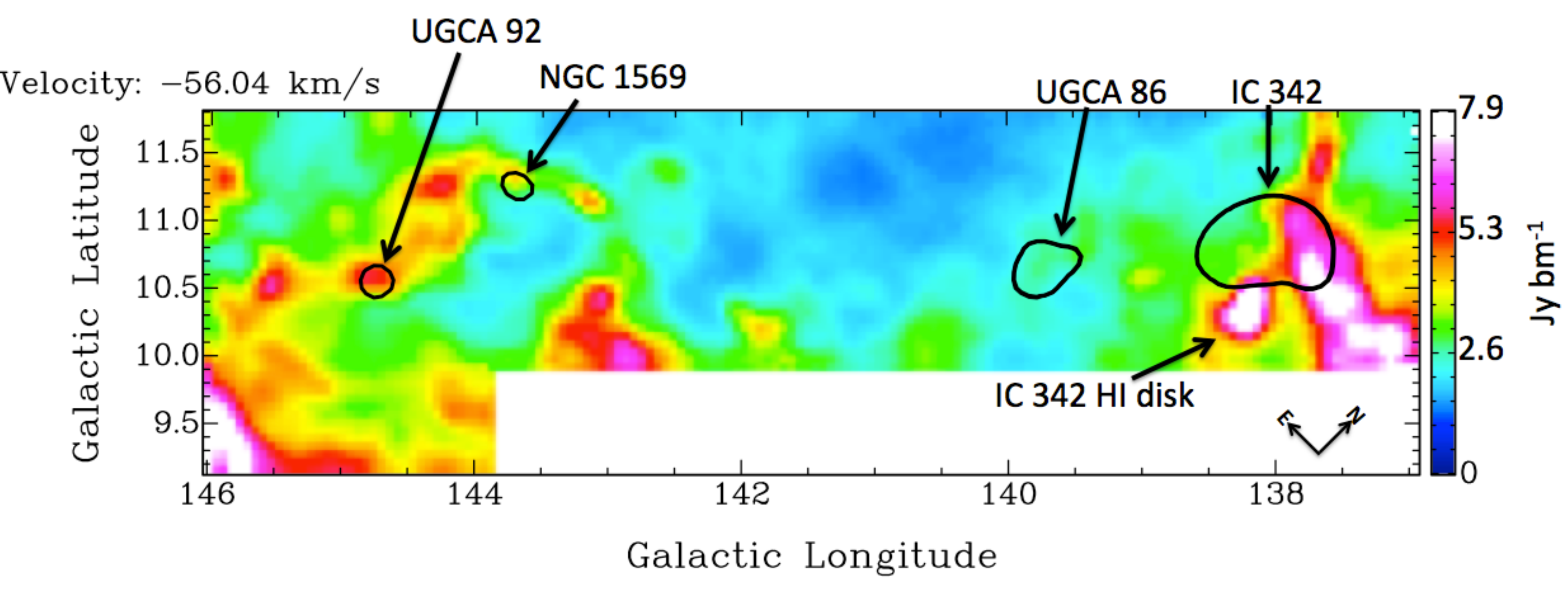}
\hfil
\includegraphics[scale=0.4]{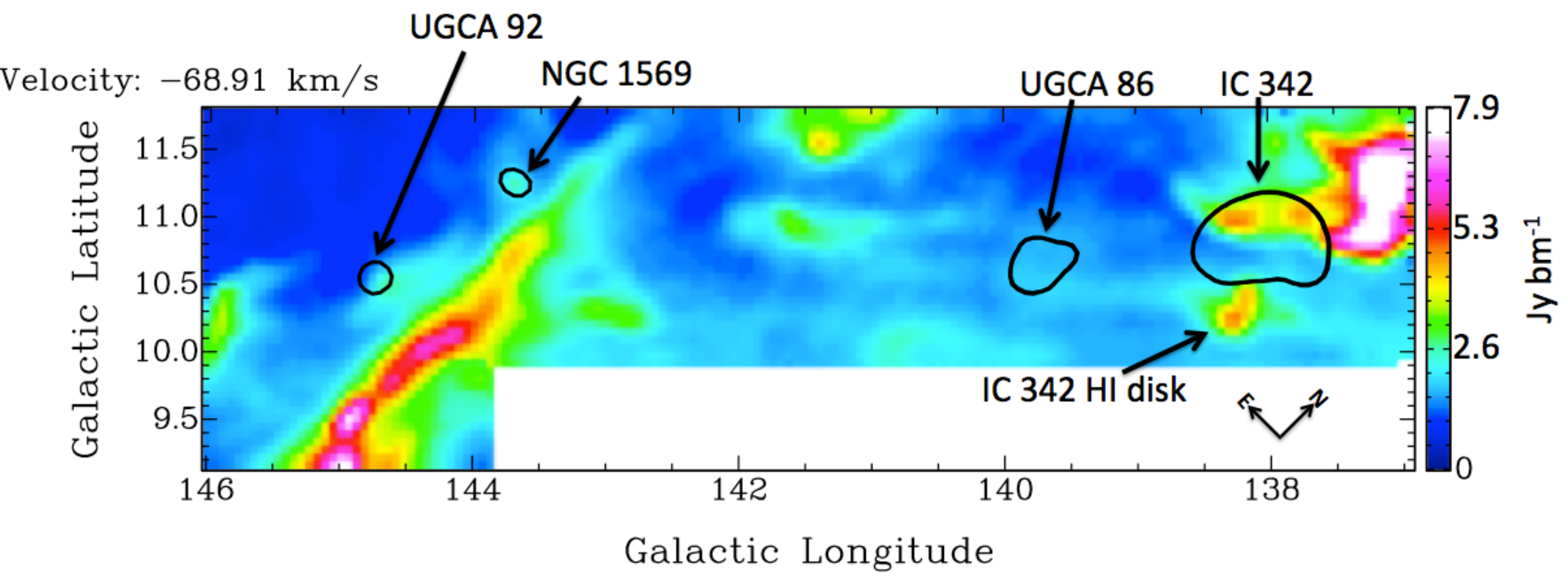}
\caption{Three single channel maps showing Milky Way emission.  These panels show the entire field of the map and the velocities of each panel are given in the upper right corner of each panel.  The intensity bar is in \jybm.  Note that the top panel shows one of the channels with the strongest Milky Way emission (note the different intensity scale), while the bottom two panels are examples of common Milky Way structures seen in the data cube.  The black contours identify the location of the four galaxies discussed in this work.  Visible in the bottom two panels is part of the \hi\ disk of IC 342 that is highly obscured by the Milky Way.  The purpose here is to show the range of Milky Way structures and how they compare to the \hdeg\ and v-shaped filaments near NGC 1569.}
\label{fig:2}
\end{figure}

\begin{figure}
\centering
\includegraphics[trim=2cm 0cm 2cm 0cm, clip=true,scale=0.4]{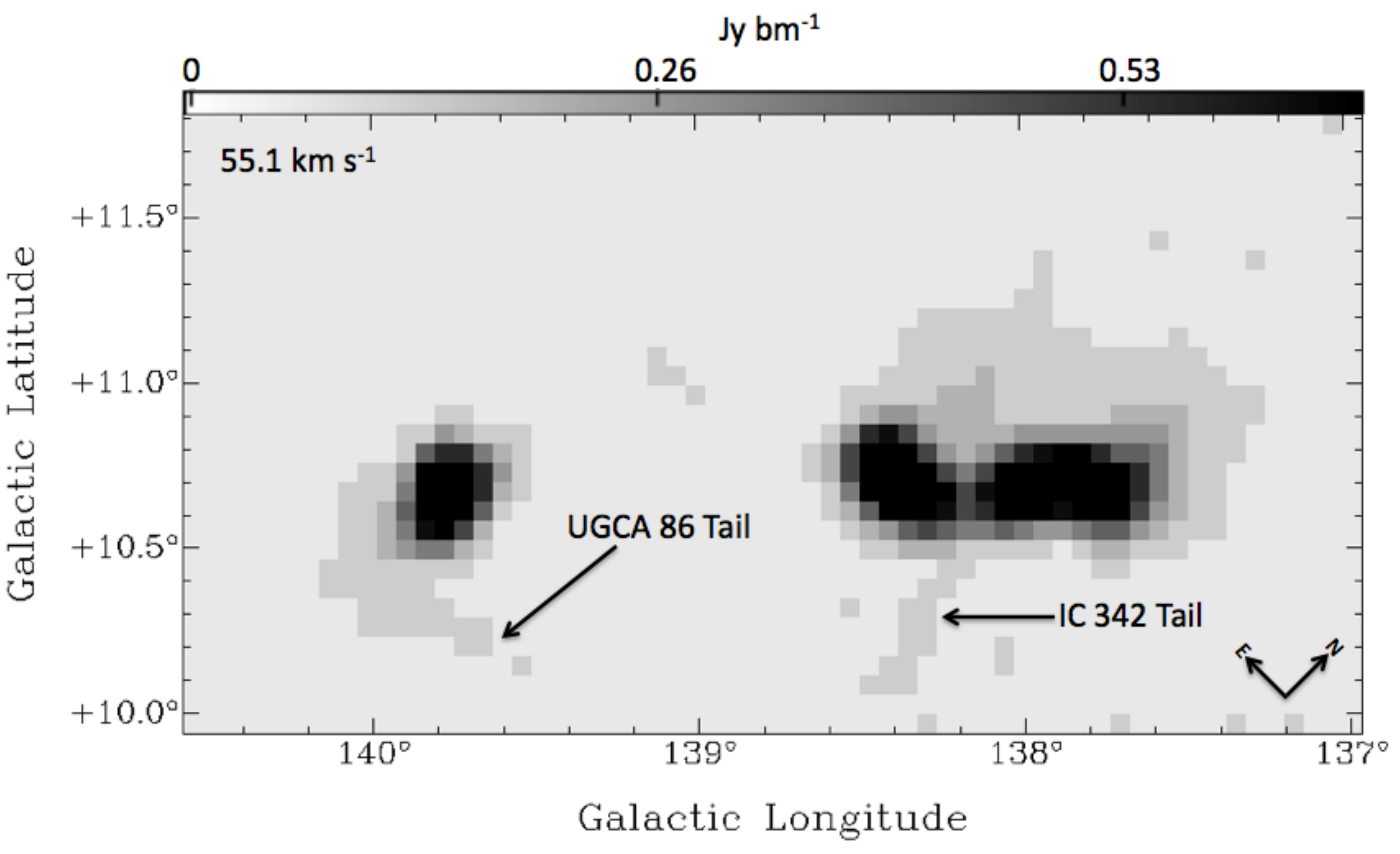}
\hfil
\includegraphics[trim=2cm 0cm 2cm 0cm, clip=true,scale=0.4]{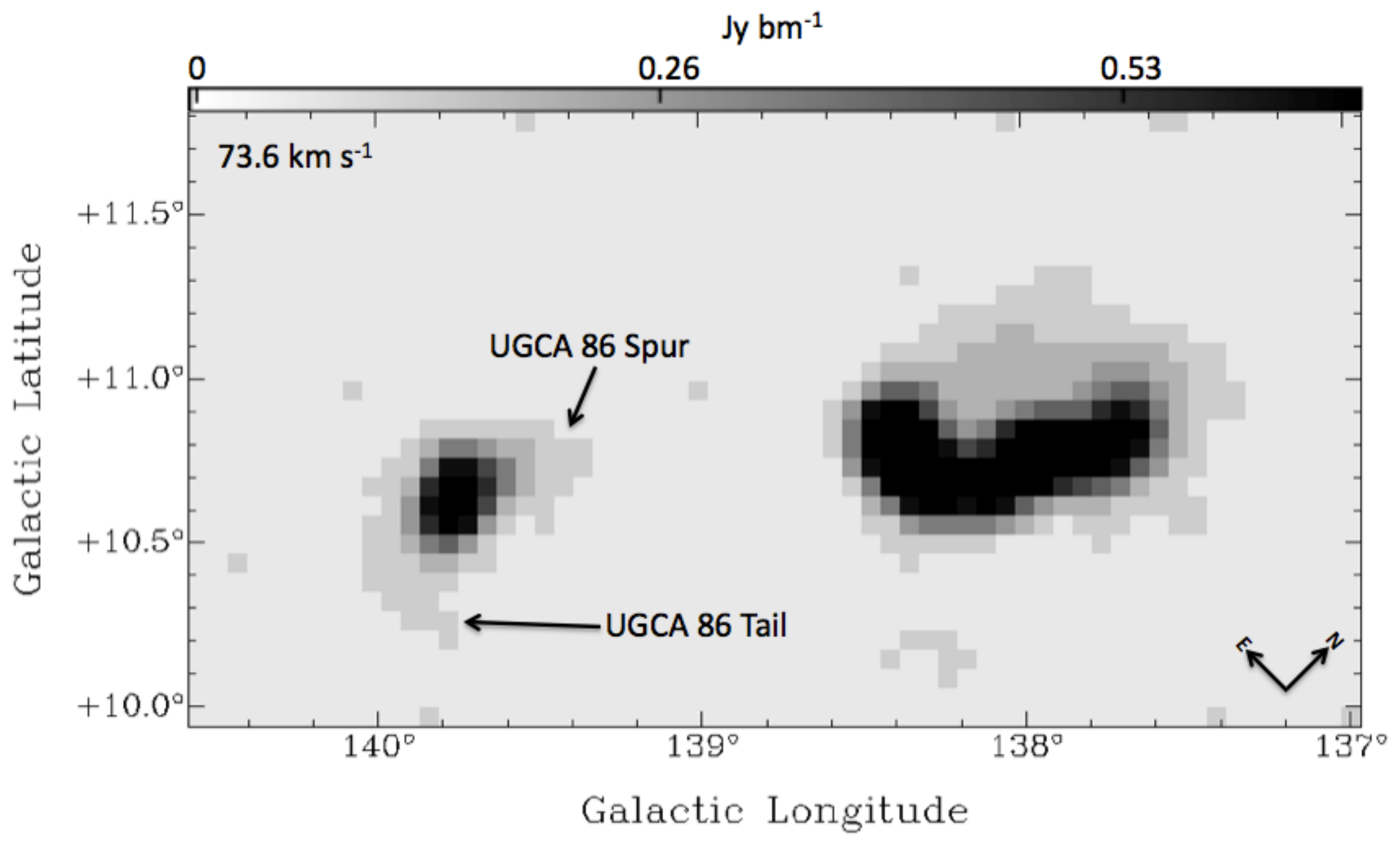}
\hfil
\includegraphics[trim=2cm 0cm 2cm 0cm, clip=true,scale=0.4]{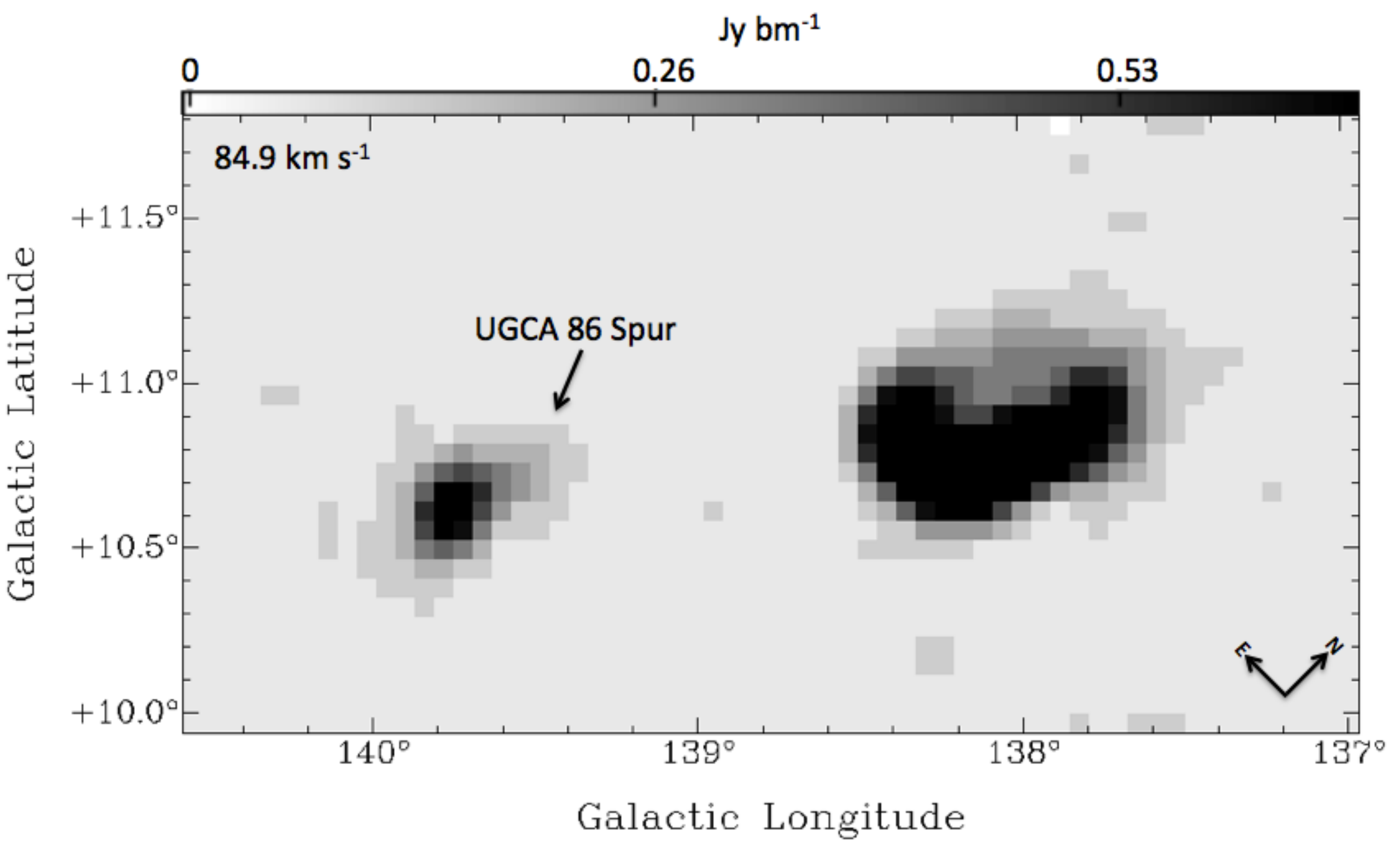}
\caption{Three single channel maps showing emission around IC 342 and UGCA 86. The \ion{H}{1} tails in IC 342 and UGCA 86 and the \hi\ spur in UGCA 86 are labeled.  The greyscale bar across the top of each panel shows the intensity in \jybm.}
\label{fig:3}
\end{figure}

\begin{figure}
\includegraphics[scale=.4]{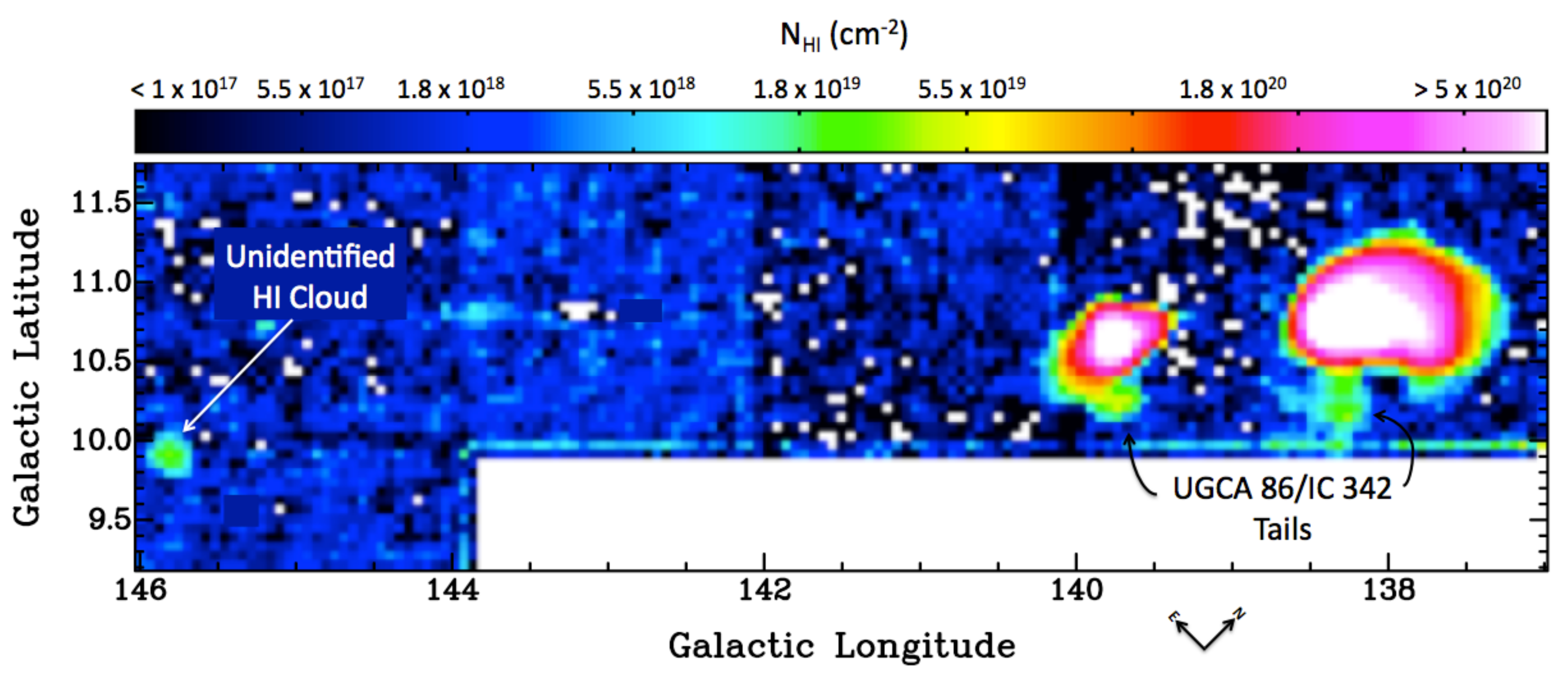} 
\caption{Integrated \hi\ intensity map made by summing over 153 \kms\ from velocity channels 36.5 to 188.8 \kms\ and shows the newly discovered tails extending from UGCA 86 and IC 342.  The tails may potentially form a continuous \hi\ bridge that extends off the southern edge of the map.  These tidal tail features have broad velocity widths and are the result of a recent interaction between UGCA 86 and IC 342.  This map also displays an unidentified \hi\ cloud near (145$\fdg7$, 9$\fdg9$).  There is no  counterpart to this cloud making its origin unknown.  The intensity scale across the top is in units of \hi\ column density.}
\label{fig:u86ic342}    
\end{figure}

\begin{figure}
\includegraphics[scale=.4]{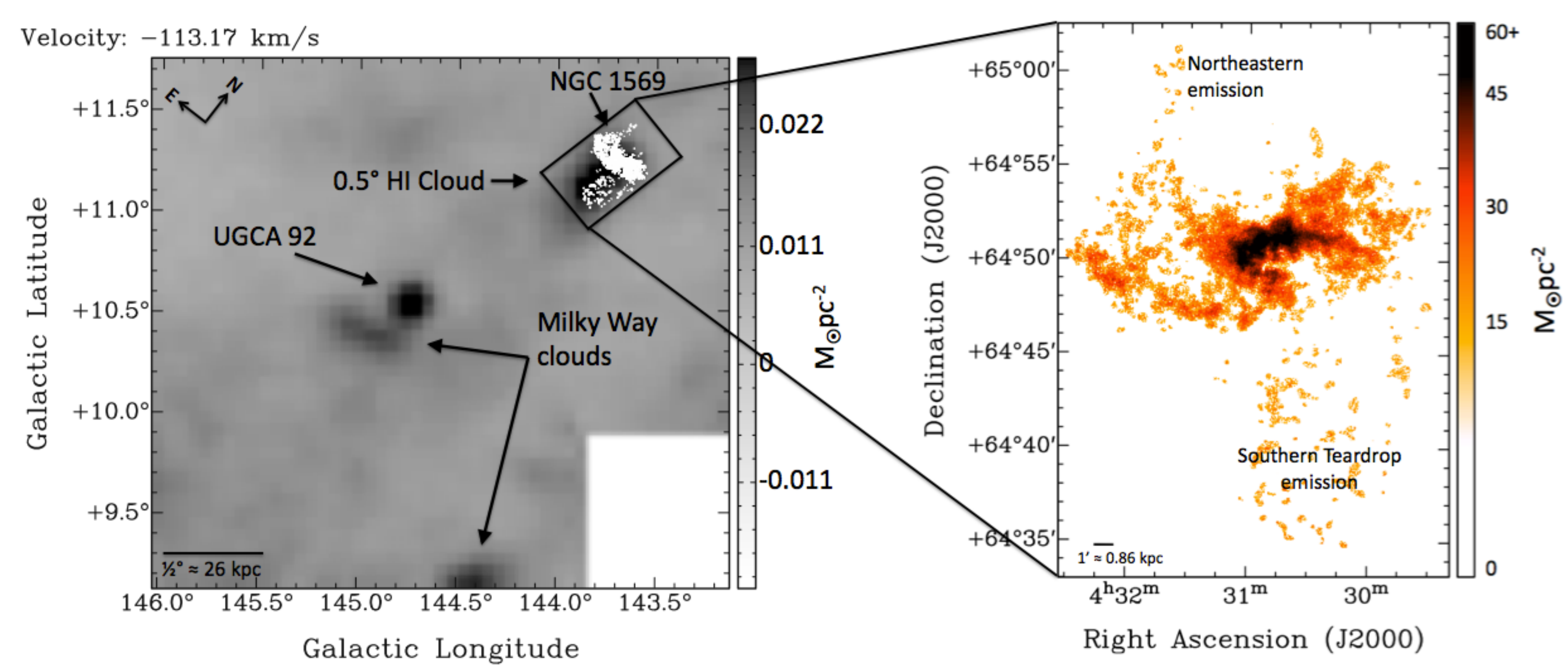} 
\caption{Individual GBT channel map showing the strongest emission from the large, \ion{H}{1} cloud over NGC 1569 extending southeast about $0\fdg5$ toward UGCA 92.  The right image is the integrated emission map from the VLA, which is Figure 5 from Paper I.  This integrated emission map was turned into contours and over plotted on the GBT channel map for reference.  For comparison, the intensity scales in the single GBT channel map (left) and the integrated flux map (right) are shown in $\rm M_\sun pc^{-2}$.  The purpose of this figure is to show the connection between the VLA map and the larger scale view of the GBT data.  The VLA contours display a diffuse teardrop shape to the south and an extension to the northeast, suggestive of a counter ``tail".}
\label{fig:4}    
\end{figure}

\begin{figure}
\centering
\includegraphics[scale=.4]{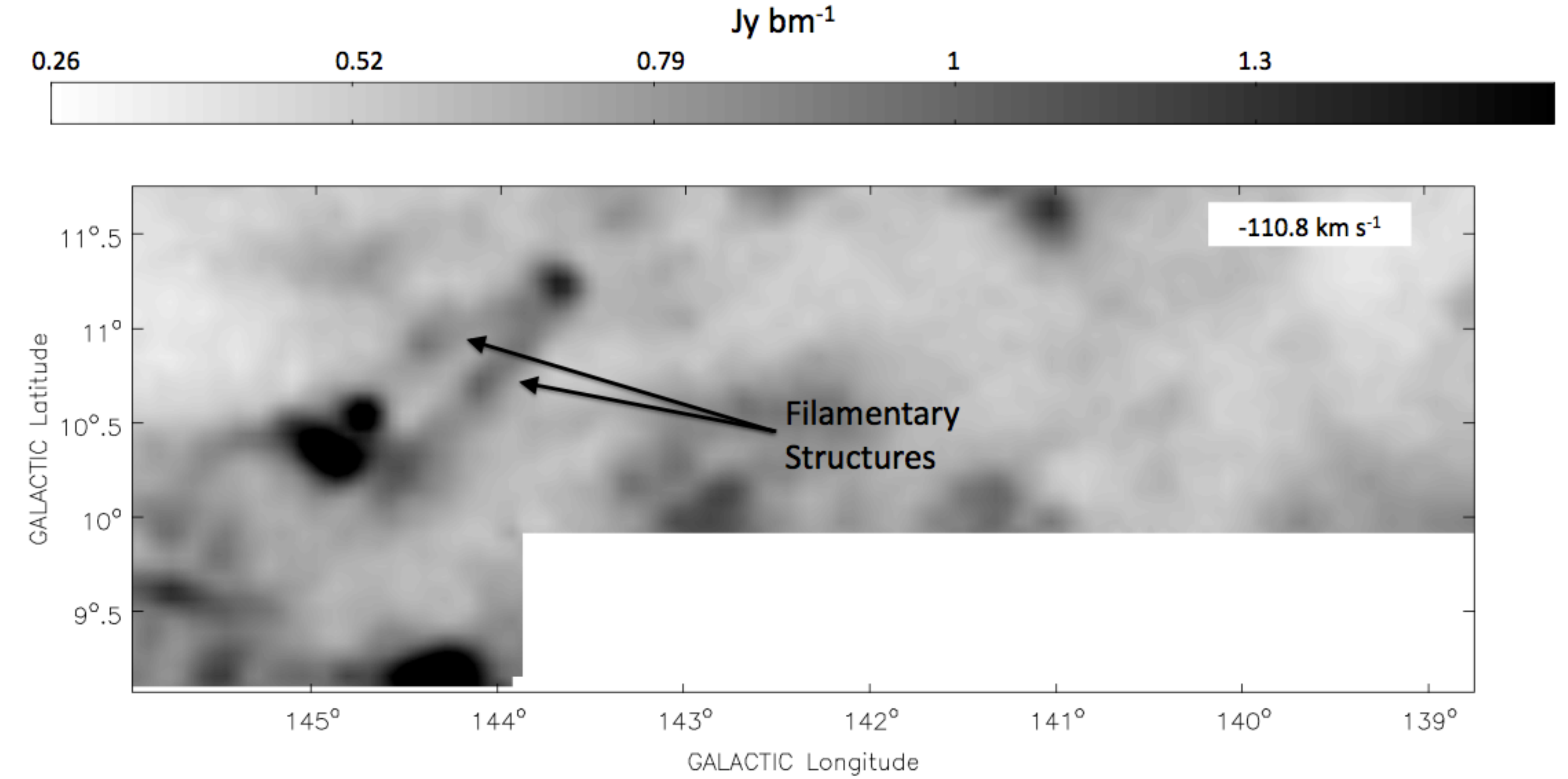} 
\caption{This individual GBT channel map has been bilineally smoothed in order to maximize the signal-to-noise.  This field displays the entire mapped area and extended filamentary emission around NGC 1569 demonstrating the suggestive extragalactic nature of these structures. The velocity is shown in the upper right corner and the intensity bar is in units of \jybm.}
\label{fig:bil}       
\end{figure}

\begin{figure}
\centering
\includegraphics[scale=.5]{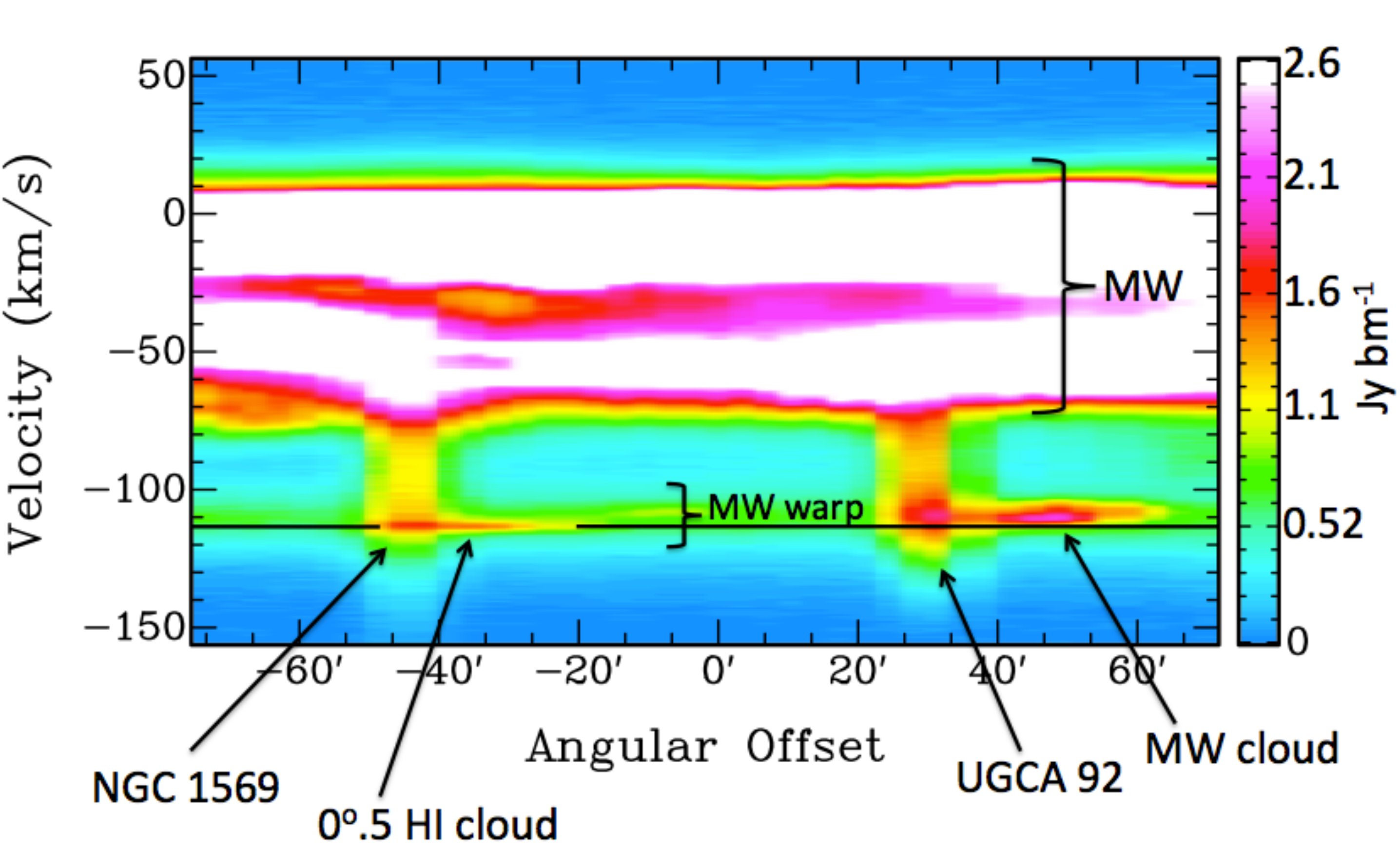}
\hfil
\includegraphics[scale=.35]{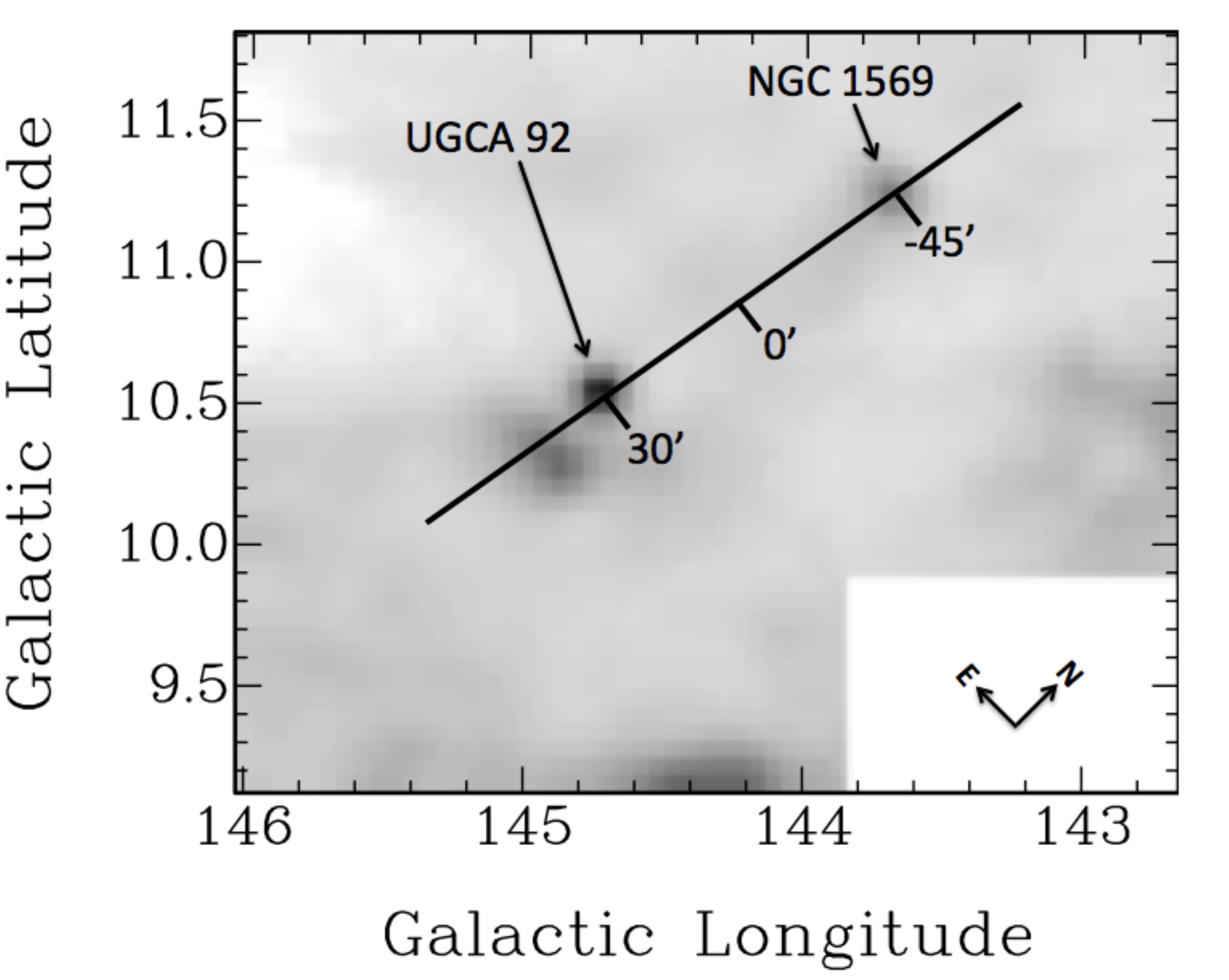}
\caption{{\it Top}:  Position-velocity diagram along the green line shown in the bottom panel that slices through NGC 1569 and UGCA 92.  The angular offset of 0$\arcmin$ corresponds to the center of the line as marked in the bottom panel. Positive direction is south, toward UGCA 92 and negative direction is north.  NGC 1569 and UGCA 92 are marked as well as the 0$\fdg5$ \ion{H}{1} cloud and the Milky Way cloud that resides to the south of UGCA 92. The horizontal black line is drawn to guide the eye along the -113 \kms\ velocity line where the \hdeg\ is strongest in emission. {\it Bottom}: Integrated intensity map of the region of the data cube containing NGC 1569 and UGCA 92.  The black line corresponds to the position of the slice used to create the position-velocity diagram in the top panel.  The galaxies are marked for reference as well as north and east. }
\label{fig:6}
\end{figure}

\begin{figure}
\subfigure[0$\fdg5$ cloud associated with NGC 1569]{\includegraphics[scale=.3]{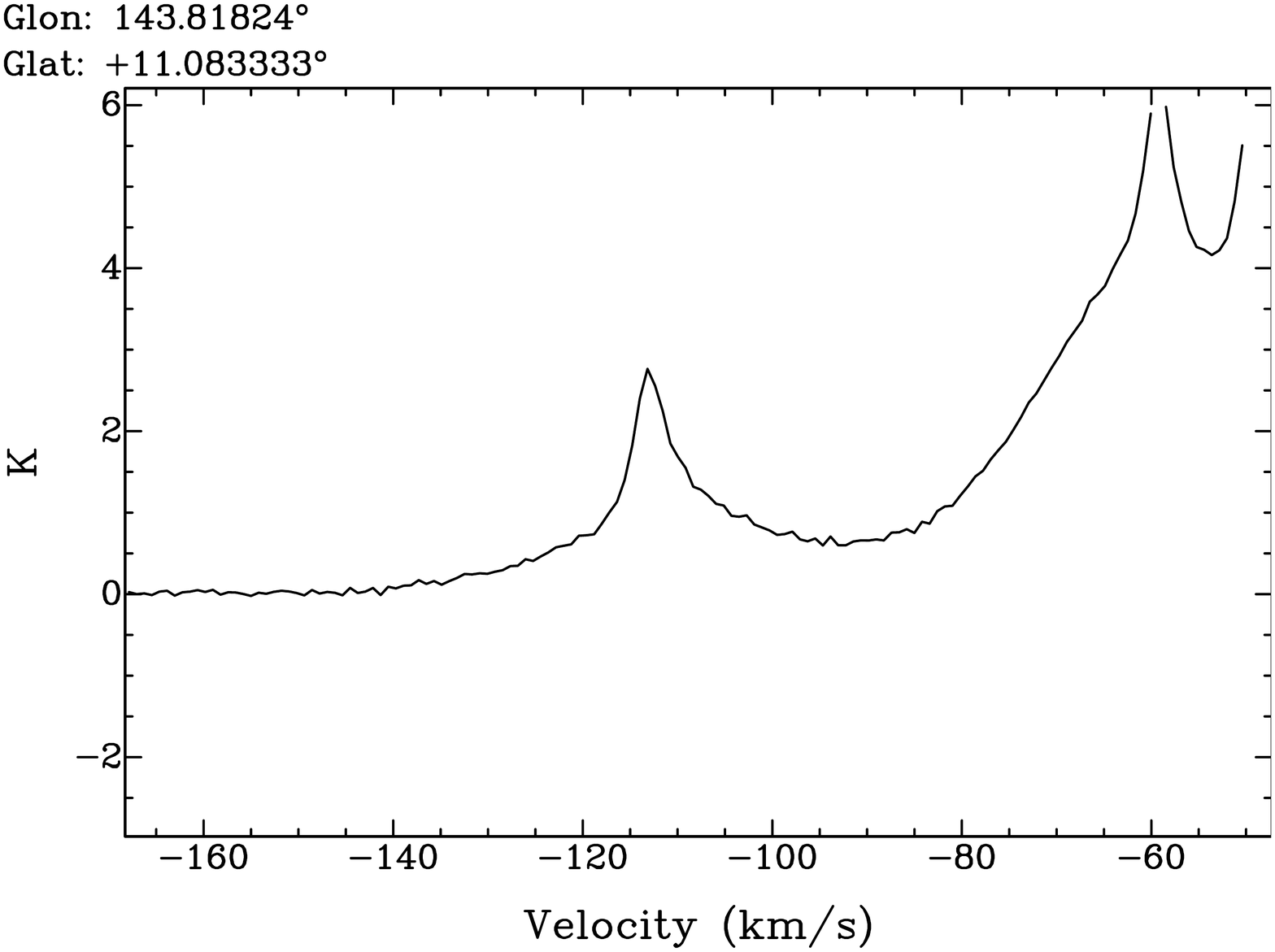}}\label{subfig:7a}
\subfigure[Milky Way cloud directly south of UGCA 92]{\includegraphics[scale=.3]{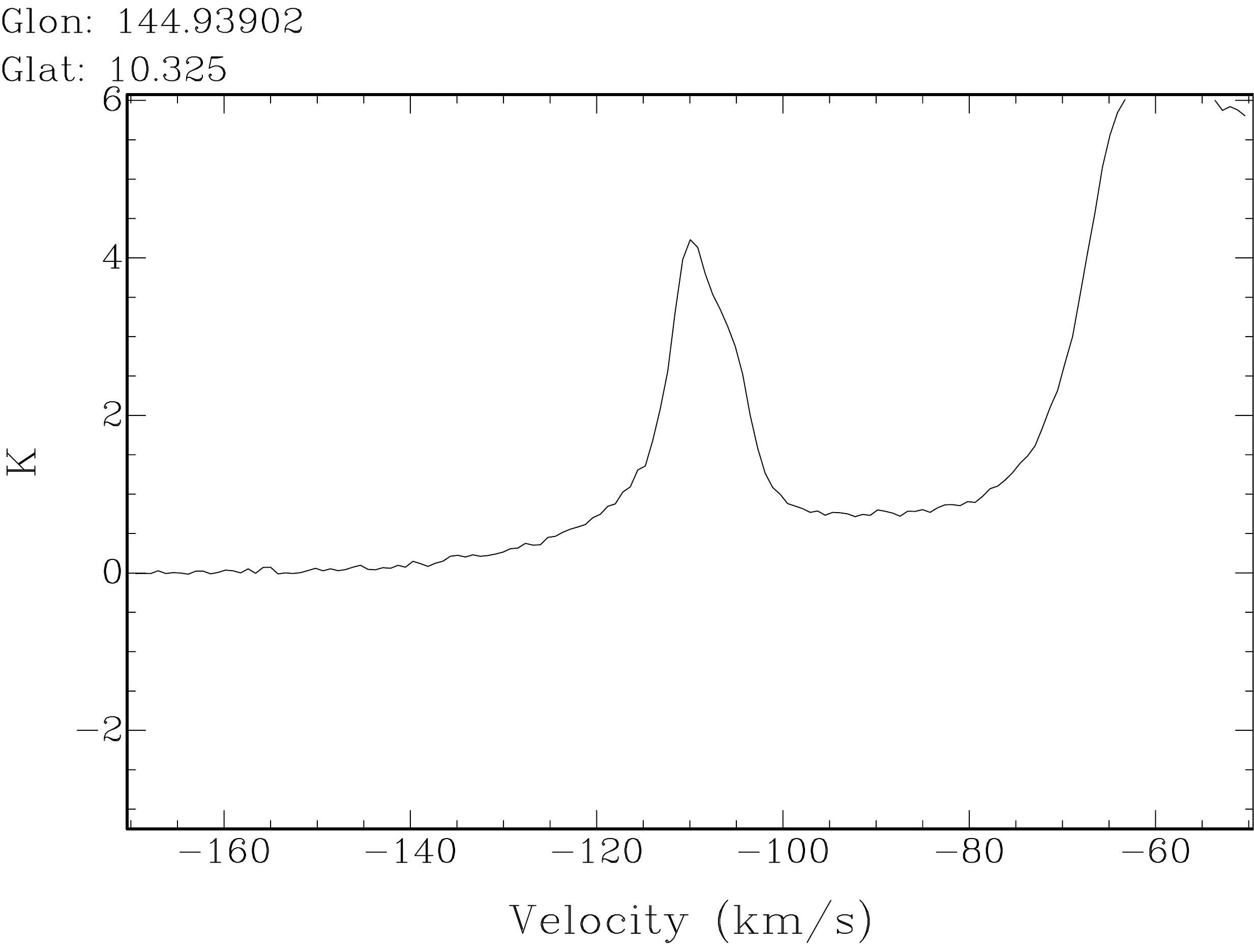}}\label{subfig:7b}
\subfigure[Milky Way cloud at bottom of map]{\includegraphics[scale=.3]{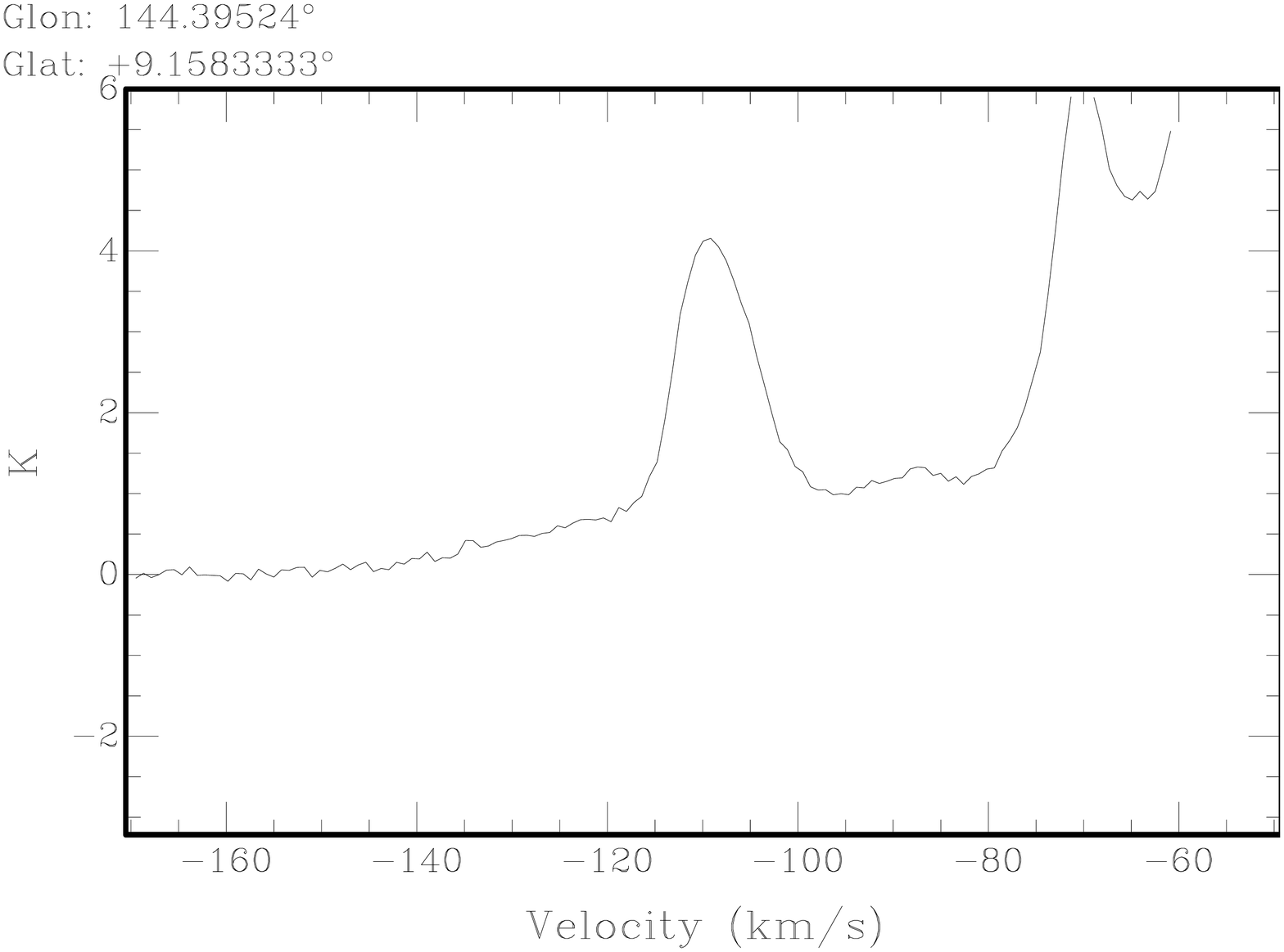}}\label{subfig:7c}
\subfigure[Center of NGC 1569]{\includegraphics[scale=.3]{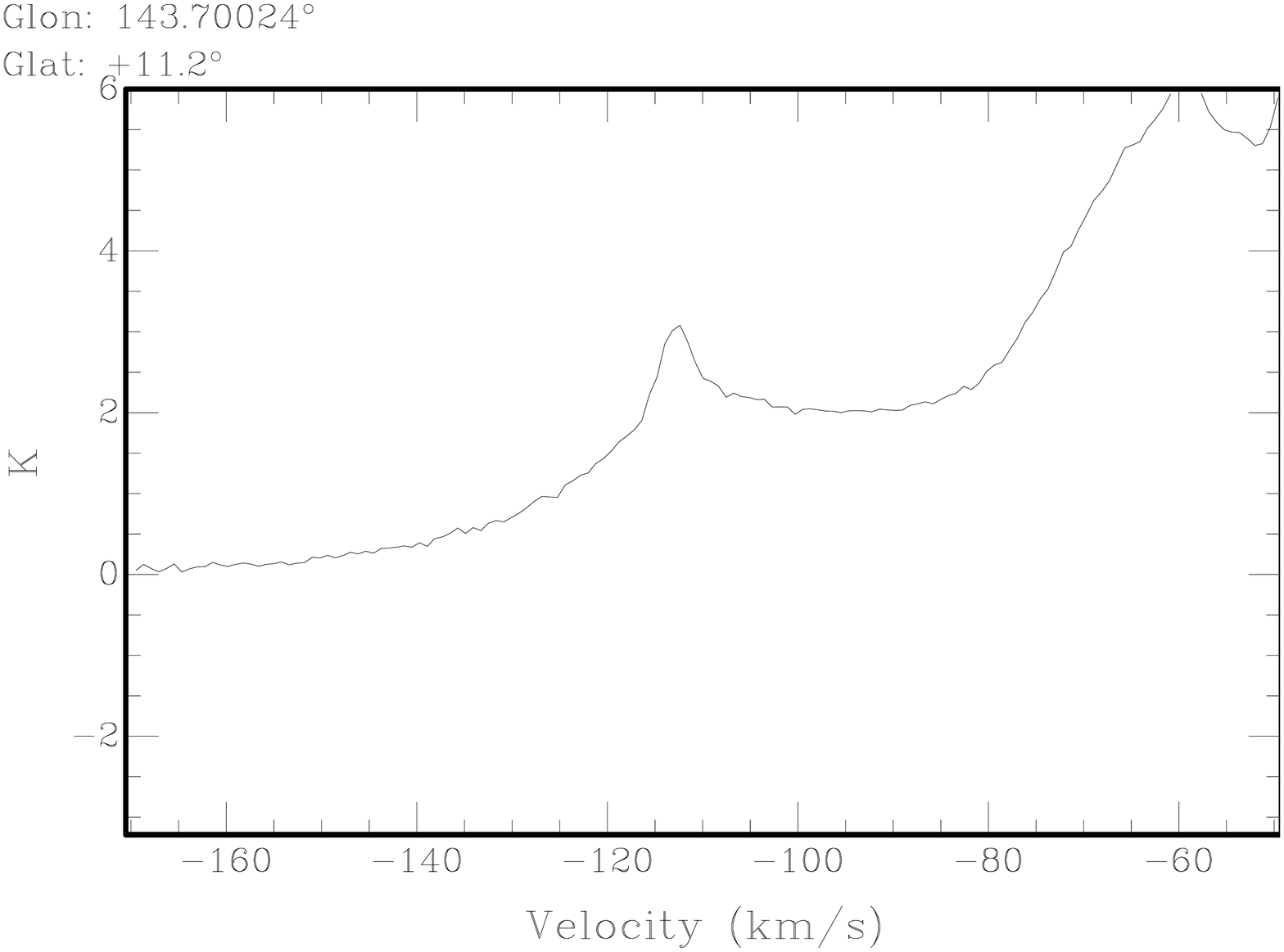}}\label{subfig:7d}
\subfigure[Center of UGCA 92]{\includegraphics[scale=.3]{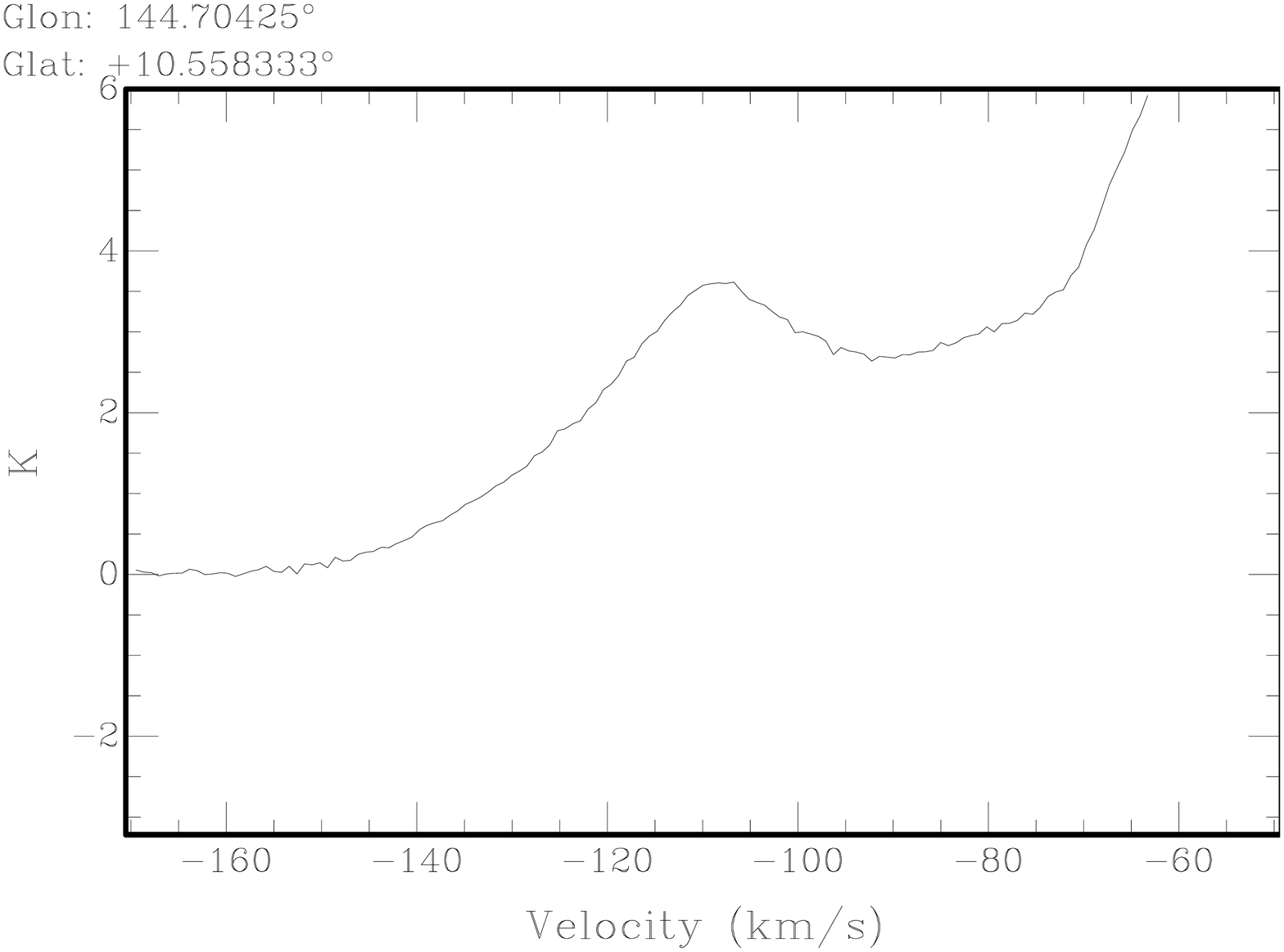}}\label{subfig:7e}
\subfigure[Single channel map showing positions of line profiles]{\includegraphics[scale=.27]{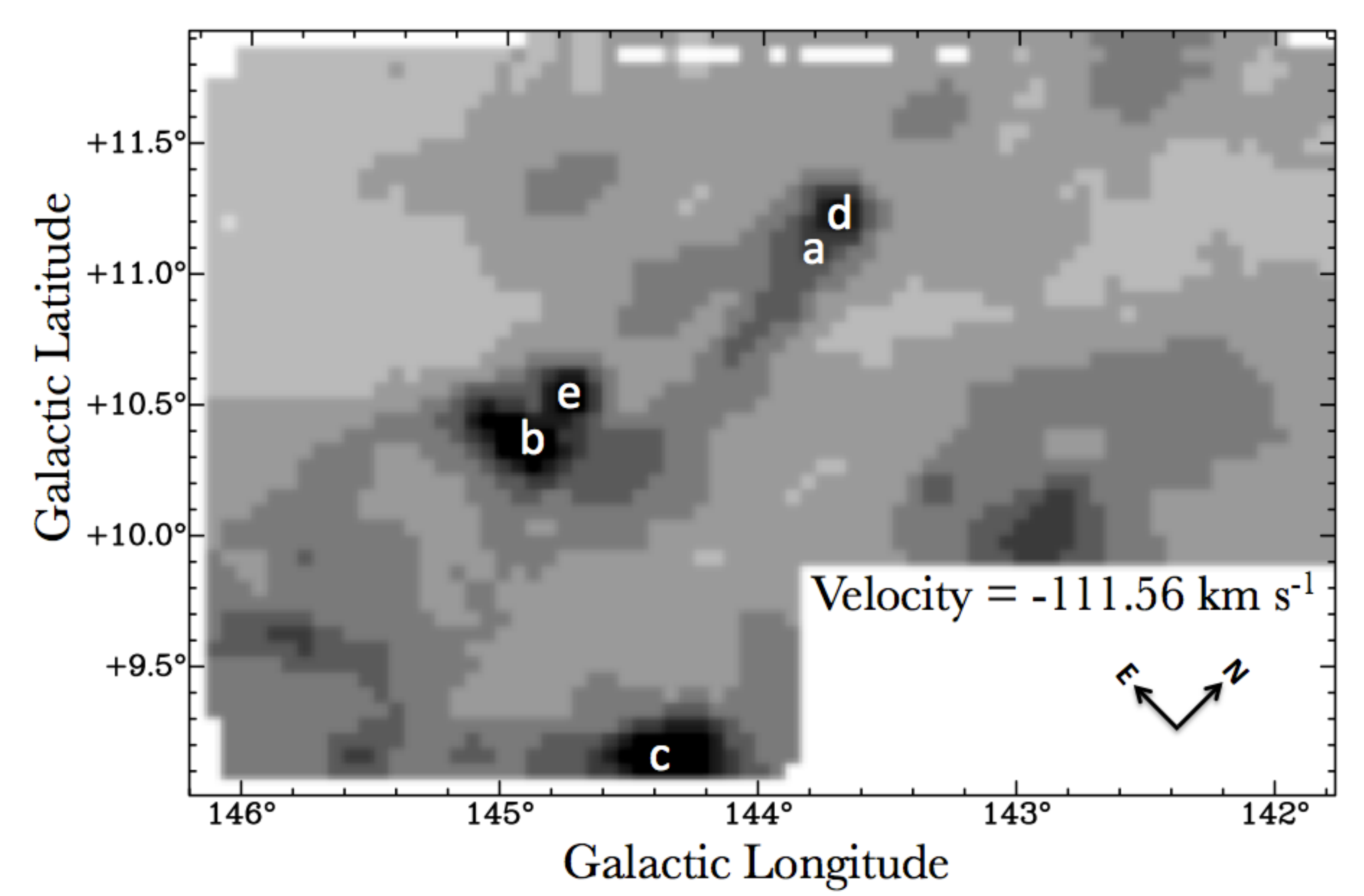}}\label{subfig:7f}
\caption{Line profiles through single pixels of the cloud centers identified in Figure \ref{fig:bil}.  The exact positions of the line profiles are given in the upper left corner of each graph.  Clouds associated with NGC 1569 are slightly red-shifted and have narrower profiles than foreground Milky Way clouds. }
\label{fig:7}
\end{figure}

\begin{figure}
\centering
\includegraphics[scale=.5]{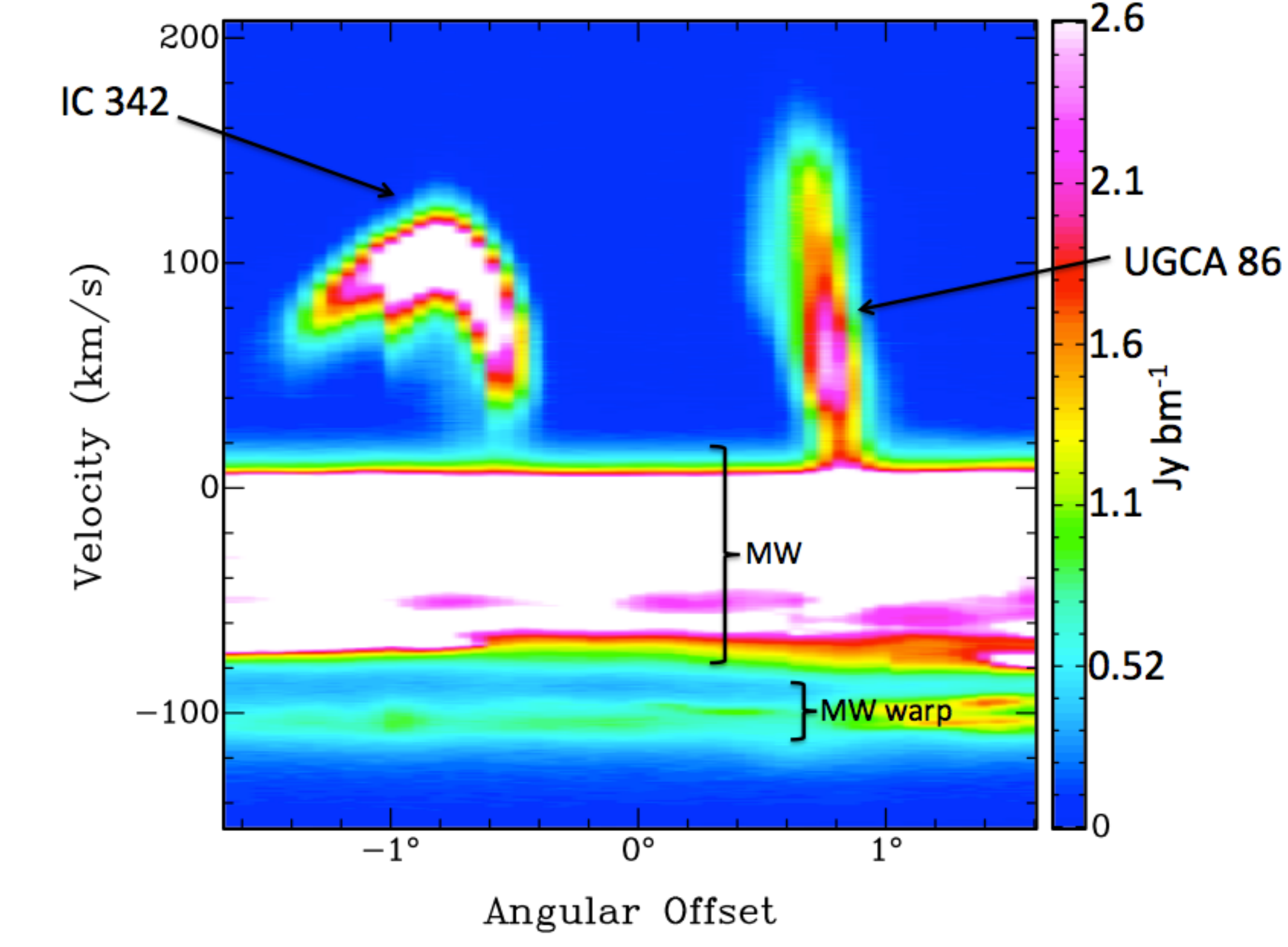}
\hfil
\includegraphics[scale=.3]{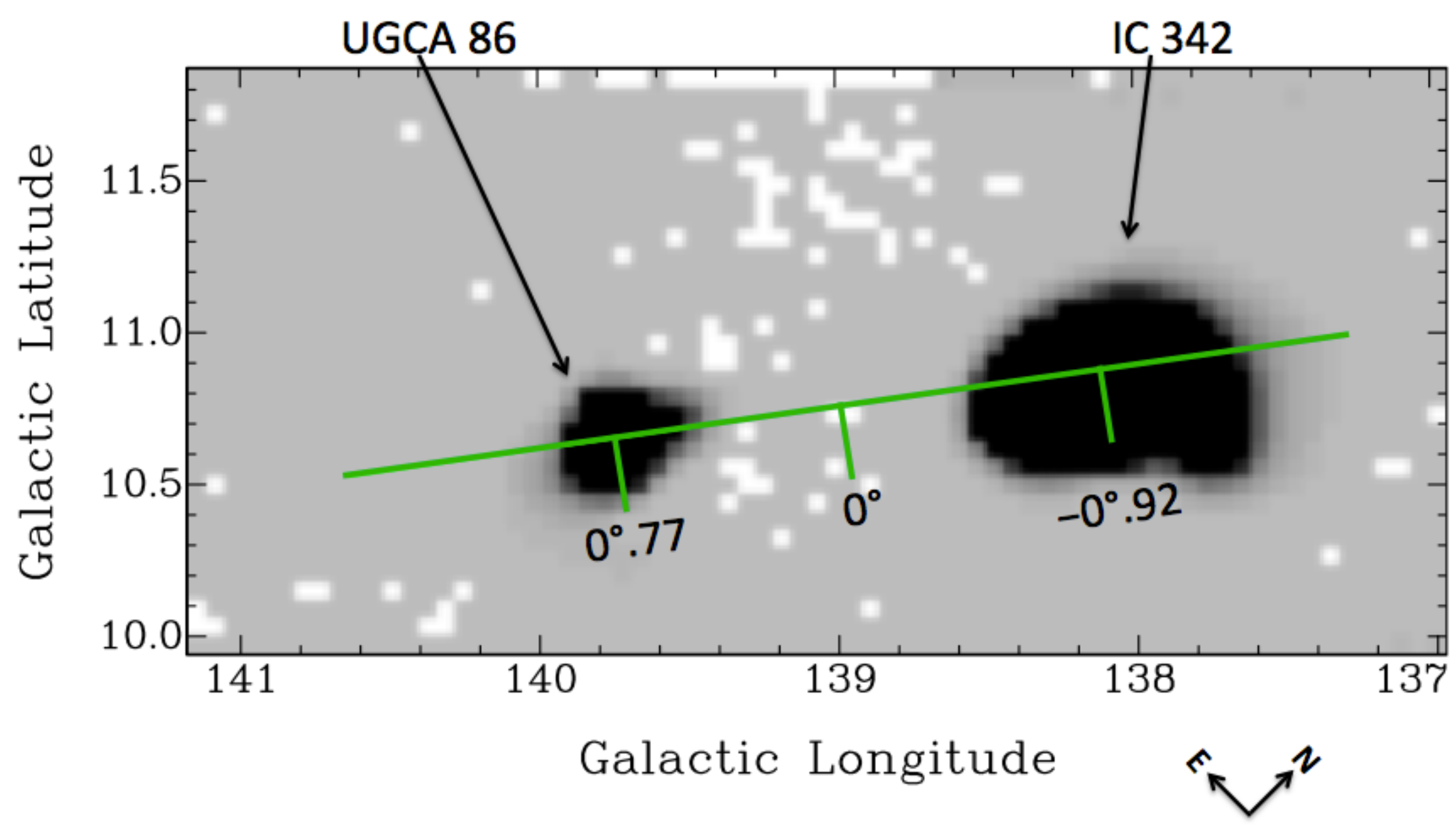}
\caption{{\it Top}:  Position-velocity diagram along the green line shown in the bottom panel that slices through the centers of IC 342 and UGCA 86.  The angular offset of 0$\arcdeg$ corresponds to the midway point between the two galaxies as indicated in the bottom panel. Positive direction is south, toward UGCA 86 and negative direction is north, toward IC 342.  IC 342 and UGCA 86 are marked as well as the Milky Way and Milky Way warp (Felix J. Lockman, private communication). {\it Bottom}: Integrated intensity map of the region including IC 342 and UGCA 86.  The green line corresponds to the position of the slice used to create the position-velocity diagram in the left panel.  The galaxies are marked for reference as well as north and east. }
\label{fig:8}
\end{figure}

\begin{figure}
\subfigure[Center of IC 342]{ \includegraphics[scale=.3]{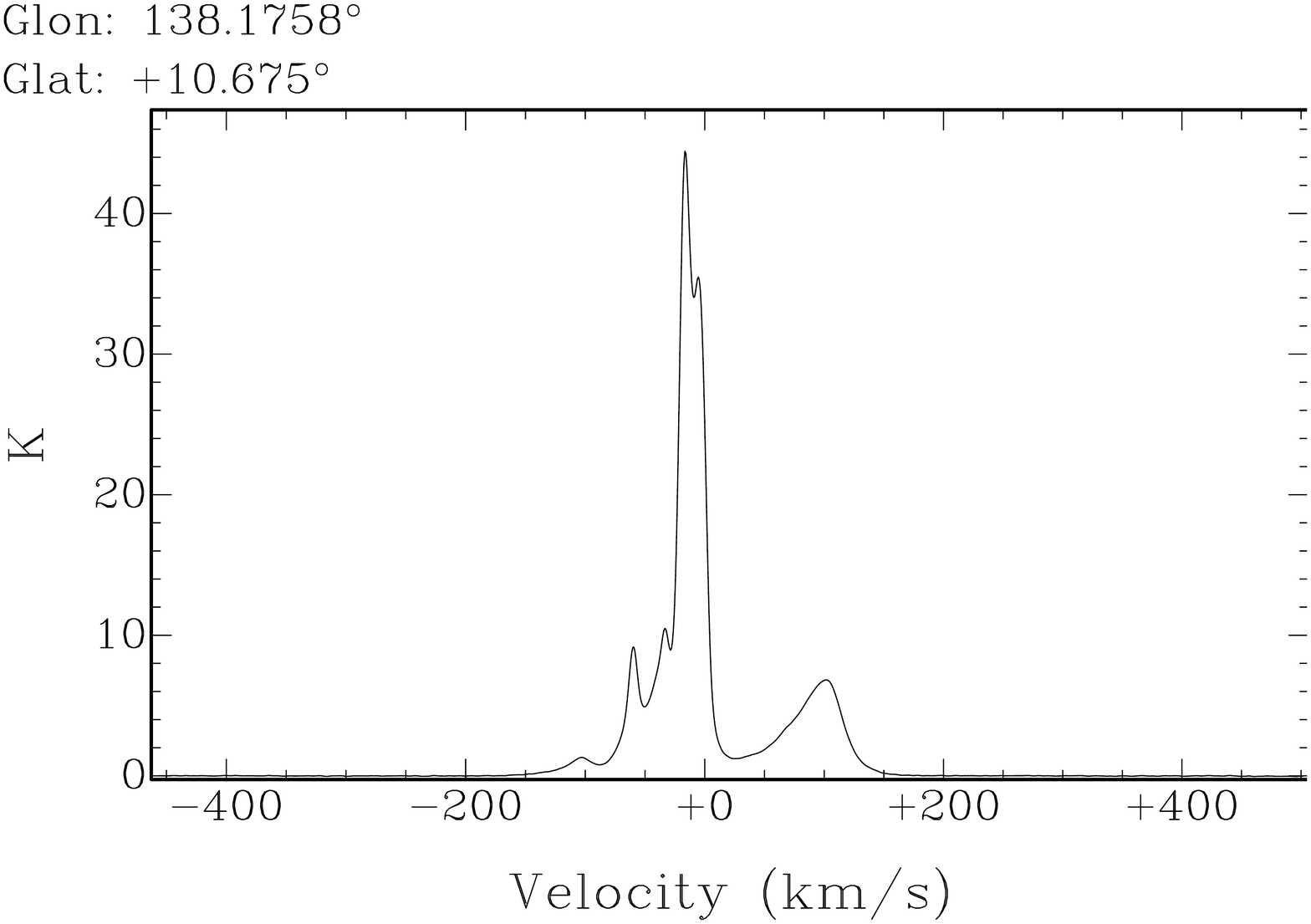}}
\subfigure[Center of UGCA 86]{\includegraphics[scale=.3]{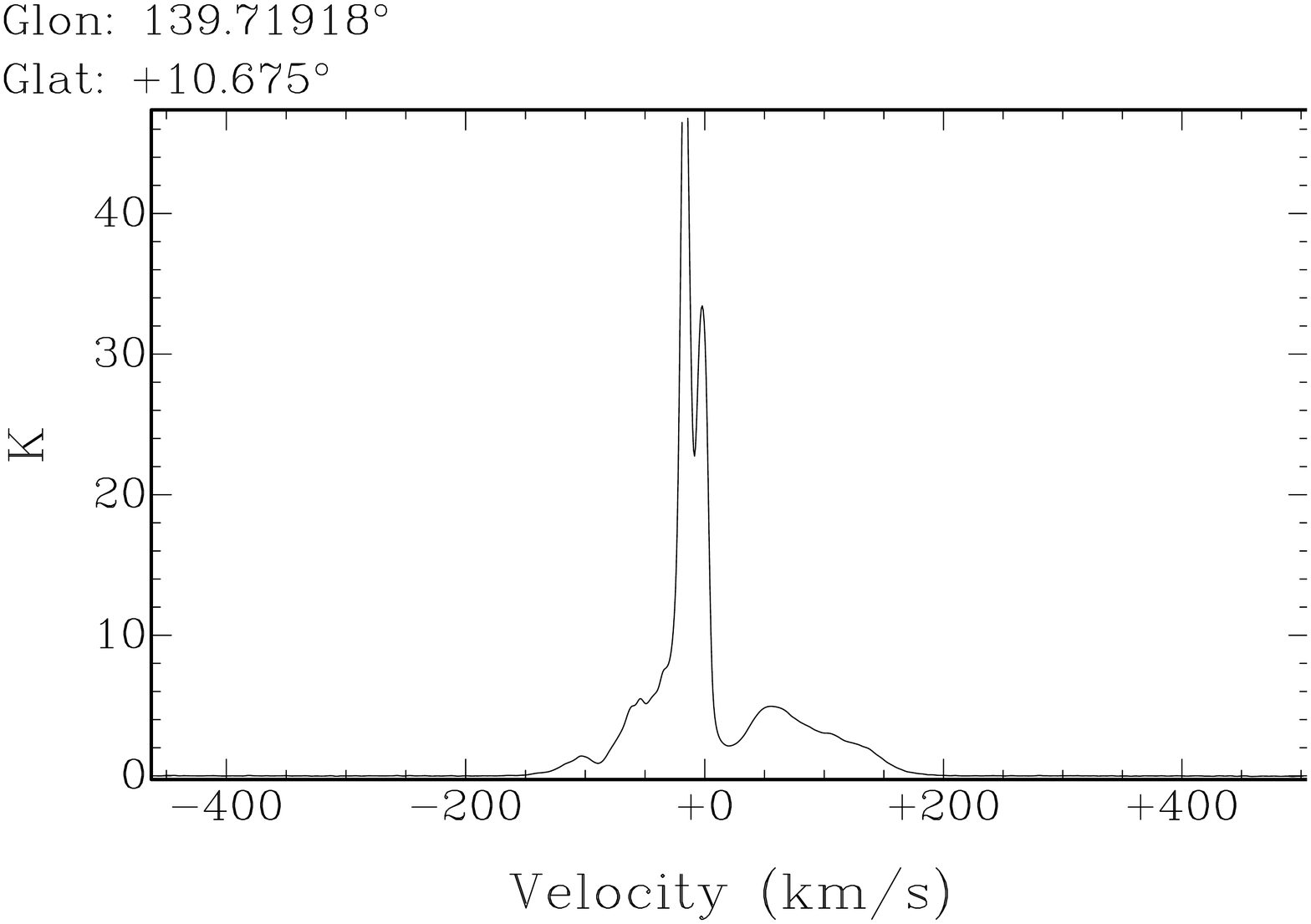}}
\caption{Line profiles through single pixels of the centers of IC 342 and UGCA 86 showing entire spectrum. The exact positions of the line profiles are given in the upper left corner of each graph. The Milky Way is at velocities less than $+$20 km s$^{-1}$. }
\label{fig:9}
\end{figure}

\clearpage

\begin{deluxetable}{lcr}
\tabletypesize{\small}
\tablenum{1}
\tablecolumns{3}
\tablewidth{0pt}
\tablecaption{GBT Observational Details}
\startdata
\cutinhead{Telescope Parameters}
Effective diameter && 100 m\\
Aperture efficiency && $\sim$ 0.71\\
Typical system temperature ($T_{\rm sys}$) && $\sim$ 20 K\\
Beam size (FWHM) && 9$\farcm1$\\
\cutinhead{Configuration Details}
Bandwidth && 12.5 MHz (2638 \kms)\\
Total number of channels && 16,384\\
Final, smoothed spectral resolution && 0.81 \kms \\
Final velocity range && -461.14 to 505.20 \kms \\
Noise per channel, $\sigma_{\rm rms}$ && $\sim$ 31 mK\\
\cutinhead{Region Mapped (Galactic coordinates)}
9$\arcdeg$ $\times$ 2$\arcdeg$ area:&&\\
\hspace{0.5cm} $l$ range && 137$\arcdeg$ to 146$\arcdeg$\\
\hspace{0.5cm} $b$ range && 10$\arcdeg$ to 12$\arcdeg$\\
Southern appendage: &&\\
\hspace{0.5cm} $l$ range && 144$\arcdeg$ to 146$\arcdeg$\\
\hspace{0.5cm} $b$ range && 9$\arcdeg$ to 10$\arcdeg$\\
\cutinhead{Observations}
Date (2010)& Hours Observed & Section of Map Observed\tablenotemark{a} \\
\hline
February 1 & 9.5&A-1, B-1, C-1, D1-1, Partial D1-2\\
February 3 & 4&Partial D1-2, A-2, Partial B-2\\
February 5 & 2.5&Partial B-2, Partial E-1\\
February 28 & 11&Partial E-1, E-2, D2-1, D2-2, \\
&&F-1, F-2, C-2, Partial A-3\\
March 8 & 3& Partial A-3, Partial B-3 \\
March 11 & 1.5&Partial B-3, Partial C-3\\
March 25 & 4& Partial C-3, Partial D-3\\
April 6&6&Partial D-3, A-4, Partial B-4\\
April 7 & 6.5 & Partial B-4, C-4, Partial D-4\\
April 8&7.5&Partial D-4, E-3, F-3, A-5\\
April 9&3&C-5, Partial D-5\\
April 13 & 2&Partial D-6\\
April 14 & 3& Partial D-6, E-4, F-4\\
April 15 & 5.5 & A-6, B-5, Partial C-6\\
April 17 & 1.5 & Partial C-6\\
April 21 & 6 & Partial C-6, B-6\\
April 23 & 4.25 & E-5, F-5, Partial E-6\\
April 24 & 2.5 & Partial E-6, F-6\\
\hline
Total Observing Hours: & 83.25 &
\enddata
\tablenotetext{a}{Refers to section identified in bottom panel of Figure \ref{fig:11}.  The number following the dash after the section letter refers to the number of passes made over that region, for example, A-4 refers to the fourth pass over region A labeled in bottom panel of Figure \ref{fig:11}.}
\label{tab:obs}
\end{deluxetable}

\clearpage

\begin{deluxetable}{ccccccc}
\tabletypesize{\small}
\tablenum{2}
\tablecolumns{7}
\tablewidth{0pt}
\tablecaption{Global Parameters for Galaxies in GBT Map}
\tablehead{\colhead{Galaxy Name} & \colhead{Type} &\colhead{$M_{\rm V}$} &  \colhead{$V_{\rm sys}$} & \colhead{Distance} & \colhead{Center}  & \colhead{Ref}\\
&&&(km s$^{-1}$)&(Mpc)&({\it l,b})&\\
(1) & (2) & (3) & (4) & (5) & (6) & (7)}
\startdata
IC 342 & SA(s)cd & -19.1 & 33 $\pm$ 4 & 3.03 $\pm$ 0.17 & (138$\fdg17$,  10$\fdg58$) & 1,3,9\\
NGC 1569 &  dIm & -18.2 & -85 & 2.96 $\pm$ 0.22 & (143$\fdg68$,  11$\fdg24$) & 2,5,7\\
UGCA 86 & SAB(s)m & -15.5 & 72 $\pm$ 5 & 2.96  $\pm$ 0.3 & (139$\fdg76$,    10$\fdg65$) & 1,6,8\\
UGCA 92 & IBm & -13.2 & -99 $\pm$ 6 & 3.01 $\pm$ 0.3 & (144$\fdg71$,   10$\fdg52$) & 1,4,8
\enddata
\tablecomments{Col.\ (1) Galaxy name; Col.\ (2) Galaxy type; Col.\ (3) Absolute $V$-band magnitude determined using distance modulus for distances in Col.\ 5 and total (apparent) $V$ magnitude.; Col.\ (4) Systemic heliocentric velocity of the galaxy with errors, when available; Col.\ (5) Distance to galaxy; Col.\ (6) Galaxy center from NASA Extragalactic Database (NED); Col.\ (7) References for type and magnitude, systemic velocity, and distance.}
\tablerefs{
Type \& Magnitude:
(1) \citet{but99};
(2) \citet{hun06}; 
Systemic Velocity:
(3) \citet{bot90};
(4) \citet{sch92};
(5) \citet{joh12};
(6) \citet{sti05};
Distance:
(7) \citet{gro12}; 
(8) \citet{kar06};
(9) \citet{fin07}
}
\label{tab:1}
\end{deluxetable}

\clearpage

\begin{deluxetable}{cccccc}
\tabletypesize{\footnotesize}
\tablenum{3}
\tablecolumns{6}
\tablewidth{0pt}
\tablecaption{Average Column Densities and \ion{H}{1} Masses for Identified Features}\tablehead{
\colhead{Object} & \colhead{Center Coords.}  & \colhead{Velocity Range} &  \colhead{$\Delta V$} & \colhead{$N_{\rm HI}$} & \colhead{$M_{\rm HI}$}\\
& ({\it l, b}) &  (km s$^{-1}$) & (km s$^{-1}$)& x10$^{18}$ cm$^{-2}$ & x10$^7$ ($M_\sun$)\\
(1) & (2) & (3) & (4) & (5) & (6)
}
\startdata
\cutinhead{IC 342} 
Galaxy & (138$\fdg17$, 10$\fdg58$) & -- & -- & -- & --\\
Tail & (138$\fdg26$, 10$\fdg24$)  & 31.7 -- 84.1 & 52.4 & 3.3 & 1.6\\
\cutinhead{NGC 1569}
Galaxy  &  (143$\fdg68$, 11$\fdg24$)  & -170.3 -- -73.7 & 96.5 & 52 & 24\\
0$\fdg5$ cloud &(143$\fdg82$, 11$\fdg08$)  & -115.6 -- -111.6 & 4.0 & 5.7 & 2.6\\
V-shaped filaments  &(144$\fdg11$, 10$\fdg85$)  & -112.4 -- -107.5 & 4.9 & 4.9 & 5.6\\
\cutinhead{UGCA 86} 
Galaxy & (139$\fdg76$, 10$\fdg65$)  & 10.0 -- 192.1 & 182.1 & 135\tablenotemark{a} & 48\tablenotemark{a}\\
Spur & (139$\fdg54$, 10$\fdg73$)  & 66.6 -- 155.8 & 89.2 & 14 & 3.1\\
Tail & (139$\fdg72$, 10$\fdg27$)  & 29.3 -- 77.6 & 48.3 & 4.2 & 2.6\\
\cutinhead{UGCA 92} 
Galaxy & (144$\fdg71$, 10$\fdg52$)  & -151.0 -- -68.9 & 82.1 & 76 & 20\\
\cutinhead{Unknown \hi\ Cloud}
Unknown \hi\ Cloud & (145$\fdg7$, 9$\fdg$9) &  33.3 -- 43.0 & 9.7 & 2.9 & --\\
\enddata
\tablenotetext{a}{The average \hi\ column density and mass for UGCA 86 are determined for the main galaxy body not including the \hi\ spur or tail and no channels that contained Milky Way emission ($<$ 10 \kms) were included.}
\tablecomments{
Col.\ (1) Galaxy/Feature name; 
Col.\ (2) Galactic coordinates of the center of the galaxy/feature; 
Col.\ (3) Velocity range over which the flux was integrated; 
Col.\ (4) Total range in velocity over which the flux was integrated; 
Col.\ (5) Average \ion{H}{1} column density of the galaxy/feature; 
Col.\ (6) Total \ion{H}{1} mass, assuming distances from column 5 in Table \ref{tab:1}.}
\label{tab:2}
\end{deluxetable}

\end{document}